%% file: CCS21.tex
\documentclass[sigconf, sigconf]{acmart}

\usepackage{tcolorbox}
\usepackage{verbatimbox}
\usepackage{balance}

\usepackage{multirow}
\usepackage{tabularx}
\def\etal{\emph{et al.}}
\def\ie{\emph{i.e.,}}
\def\eg{\emph{e.g.,}}

\AtBeginDocument{%
  \providecommand\BibTeX{{%
    \normalfont B\kern-0.5em{\scshape i\kern-0.25em b}\kern-0.8em\TeX}}}

\setcopyright{acmlicensed}
\begin{document}
\pagestyle{plain}
\copyrightyear{} 
\acmYear{} 
\acmConference[]{}{}{}
\acmPrice{}
\acmDOI{}
\acmISBN{}

\title{A Tandem Framework Balancing Privacy and Security for Voice User Interfaces}

\author{Ranya Aloufi, Hamed Haddadi, David Boyle}
\affiliation{%
  \institution{Imperial College London}
  \streetaddress{}
  \city{}
  \state{}
  \country{}
  \postcode{}
}

\begin{abstract}
Speech synthesis, voice cloning, and voice conversion techniques present severe privacy and security threats to users of voice user interfaces (VUIs). These techniques transform one or more elements of a speech signal,~\eg, identity, emotion, or accent, while preserving linguistic information. Adversaries may use advanced transformation tools to trigger a spoofing attack using fraudulent biometrics for a legitimate speaker. Conversely, such techniques have been used to generate privacy-transformed speech by suppressing personally identifiable attributes in the voice signals, achieving anonymization. Prior works have studied the security and privacy vectors in parallel, and thus it raises alarm that if a benign user can achieve privacy by a transformation, it also means that a malicious user can break security by bypassing the anti-spoofing mechanism. 

In this paper, we take a step towards balancing two seemingly conflicting requirements: security and privacy. It remains unclear what the vulnerabilities in one domain imply for the other, and what dynamic interactions exist between them. A better understanding of these aspects is crucial for assessing and mitigating vulnerabilities inherent with VUIs and building effective defenses. In this paper, (i) we investigate the applicability of the current voice anonymization methods by deploying a tandem framework that jointly combines anti-spoofing and authentication models, and evaluate the performance of these methods; (ii) examining analytical and empirical evidence, we reveal a duality between the two mechanisms as they offer different ways to achieve the same objective, and we show that leveraging one vector significantly amplifies the effectiveness of the other; (iii) we demonstrate that to effectively defend from potential attacks against VUIs, it is necessary to investigate the attacks from multiple complementary perspectives (\ie~security and privacy) and carefully account for the effects of deploying countermeasures, pointing to several promising research directions.


\end{abstract}

\ccsdesc[500]{Security and privacy~Privacy protections}

\keywords{Voice Conversion, Automatic Speaker Verification, Anti-spoofing, Privacy \& Security}

\maketitle

\input{introduction}
\input{background}
\input{taxonomy}
\input{experiments}
\input{discussion}
\input{related_work}
\input{conclusion}

\bibliographystyle{ACM-Reference-Format}
\balance
\bibliography{ccs21}

\end{document}

%% file: introduction.tex
\section{Introduction}

Voice User Interfaces (VUIs) for conversational agents are now common in many services such as banking, call centers, and medical services, in addition to voice assistants (VA) like Amazon Alexa and Google Assistant. Often these services rely on verifying the user through speaker recognition and then using speech recognition techniques to understand a spoken language. Increased reliance on VUI services exposes users to an increasing number of threats to their privacy and security. Speech recordings are a rich source of personal information~\cite{schuller1988emotion}, and the degree of privacy-sensitive information (\eg~emotion, sex, accent, and ethnicity) captured in these recordings extends beyond what is said (\ie~linguistic) and who says it (\ie~paralinguistic). Security and privacy concerns arise from the potential interception and misuse of this sensitive information sharing through automatic speech processing technology.

In the security domain, we can consider an adversary which aims to fool the target model~\cite{papernot2016transferability}. Evasion attacks, also known as adversarial examples, add imperceptible perturbation to the input sample to result in the incorrect prediction of the target models (\eg~automatic speech recognition (ASR) and automatic speaker verification (ASV))~\cite{9383529, carlini2018audio, abdullah2019practical, abdullah2019hear, cisse2017houdini, yuan2018commandersong, taori2019targeted, qin2019imperceptible, schonherr2018adversarial, abdoli2019universal, alzantot2018did}. A spoofing attack (\ie~replay, synthesis, and voice conversion attacks) is a technique where the imposter speaker’s speech is converted to desired speaker’s speech using signal processing approaches that cause false acceptances to authentication systems~\cite{wu2014voice}. In response to these potential attacks, the countermeasures for adversarial/spoofing attacks have been proposed to secure target models against these attacks ~\cite {wang2020asvspoof, 9053086, hemavathi2021voice}.

In the privacy domain, the adversary aims to obtain private information about the training data or to obtain the model itself~\cite{8835245}. Attacks targeting data privacy include, for example, an attacker aiming to determine if the voice of a certain individual was used for training a speaker identification system. In response to these potential attacks, privacy-preserving defenses have been designed to prevent privacy leakage of the raw data. These defenses fall between anonymization and cryptography~\cite{tomashenko2020introducing}. For example, anonymization aims to make the speech input unlinkable, i.e., ensure that no utterance can be linked to its original speaker by altering a raw signal and mapping the identifiable personal characteristics of a given speaker to another identity~\cite{9053868}. Various studies have proposed anonymization methods based on noise addition~\cite{tomashenko2020introducing}, voice conversion~\cite{9053868, 255304, srivastava2020design}, speech synthesis~\cite{Hidebehind_2018}, and adversarial learning ~\cite{Srivastava_2019}, considering the speaker identity~\cite{tomashenko2020introducing} or emotion~\cite{emotionless_2019} as a sensitive attributes.  

Voice conversion (VC) is a technique used to convert paralinguistic information such as gender, speaker identity, and emotions while keeping the linguistic information of a source speech. These technologies were made much more powerful by incorporating deep learning mechanisms. Recently, VC technology has become a key technology in designing privacy-preserving voice analytics solutions to produce convincing mimicry of specific target speaker voices. For example, Srivastava~\etal~\cite{srivastava2020design} designed an anonymization scheme that converts any input speech into that of a random pseudo-speaker. Ho~\etal~\cite{9367139} propose a speaker identity-controllable framework based on VC technology to mimic voice while continuously controlling speaker individuality. On the contrary, such techniques can enable fooling (spoofing) unprotected speaker authentication systems and therefore might prompt various potential security implications. With the generated spoofed recordings, for instance, an adversary might attack the voice assistant, making it fraudulently respond to identity-based service requests; an insider might attack the VoicePrint-based security system to gain illegitimate access and gain sensitive information; an imposter might call a bank’s contact center by making himself recognized as the victim. Thus, it raises alarm about the feasibility of achieving privacy by applying voice transformation (\ie~anonymization) regarding its security threat in real-time practical applications such as smart-assistance systems.

With the increasing use of automatic speaker verification (ASV) in security-sensitive domains (\eg~forensics identification and smart-home), ASV is becoming a new target for attackers. It has been shown that ASV systems can be vulnerable to fooling/spoofing, also referred to as presentation attacks~\cite{wang2020asvspoof}, since these systems generally are not yet efficient in recognizing voice modifications/variations (\ie~adversarial examples, noisy voice samples, mismatch conditions between enrolling and trails recordings)~\cite{dehak2010front}. VC technology could be also misused for attacking these systems, and thus spoofing countermeasures (CM) have been proposed and adopted to protect ASV systems. CMs are designed to learn the distinguishing artifacts present in spoofed audio produced by VC from human speech. Spoofing refers to falsifying a speech signal as system input for feature extraction and verification, the objective of which is to improve the reliability of biometric systems by preventing fraudulent access. While the ASV system should reject a zero-effort impostor (\ie~false attempt), the CMs should detect a valid trial (\ie~genuine speech). 

While security, privacy, and data protection are often studied independently, there is little understanding on their fundamental interconnections, and the complexity of their relation has not been fully explored. Specifically, in the speech domain, prior work has intensively studied the two domains separately~\cite{carlini2018audio, abdullah2019practical, emotionless_2019}. Thus, it remains unclear what the vulnerability to one domain implies for the other. Revealing such implications is important for developing effective defenses where security and privacy can be co-engineered. It is unclear how the two vectors interact with each other and how their interactions may influence attack dynamics against VUIs systems. Understanding such interactions is critical for building effective defenses. For example, in voice assistance systems, the users need to be verified first using voice-based authentication to gain access to further services (\eg~understanding the user command and responding based on it), assuming that anonymization mechanism is detected to protect user privacy (\ie~hiding sensitive speaker-related information), resulting in modified/synthesized voices that can affect the authentication functionality or be blocked by CMs that may detect it as a spoofed signal. Further, the adversary may exploit such an anonymization tool to mislead the authentication operation. Finally, studying potential attack vectors within a unified framework is essential for assessing and mitigating the broad vulnerabilities of VUIs deployed in practice, in which multiple attacks may be launched simultaneously. 
In this paper, we seek to answer the following research questions.\\
\textbf{RQ1 –} What are the fundamental connections between voice spoofing and voice anonymization?\\
\textbf{RQ2 –} What are the implications of such mechanisms (\eg~speech synthesis, voice cloning, and voice conversion) for an adversary to optimize attack strategies against VUI-enabled services, and for benign users to protect their privacy?\\
\textbf{RQ3 –} What are the potential countermeasures to maintain secure and private VUI-based systems?\\

\textbf{Our Contribution}. In this work we present a step towards answering the key questions above. Answering these key questions is crucial for assessing and mitigating the broad vulnerabilities of VUIs deployed in realistic settings.\\

\emph{RA1} – We use a tandem framework that jointly investigates ASV and CM models performance against two vectors of attacks generated by voice transformation and anonymization mechanisms. With this framework, we show that there exists an intricate duality between the two mechanisms. Specifically, they offer different ways to achieve the same objective.\\

\emph{RA2} – Through empirical studies on benchmark datasets and using both spoofing countermeasures and anonymization techniques, we reveal that the anonymized voices are detected as spoofed attacks, intuitively, leading to confusingly questioning its effectiveness in obtaining privacy-transformed utterances to meet the anonymization purposes. We also provide analytical justification for such effects under a different setting.~\footnote{Code and research artefacts.~\url{https://github.com/RanyaJumah/EDGY/tree/master/Balancing_Privacy&Security_for_VUI}}\\
\emph{RA3} – Finally, we demonstrate that to effectively defend against attacks, it is necessary to consider attacks from multiple complementary perspectives (\ie~security and privacy) and carefully account for the effects in applying the mitigating solutions.\\

To our best knowledge, this work represents the first systematic study of voice spoofing (\ie~for security deceiving) and anonymization (\ie~for privacy protection) within a unified framework. We believe our findings deepen understanding of the vulnerabilities of VUIs in practical settings and shed light on how to develop more effective, secure \emph{and} private solutions.

\begin{figure}[t!]
  \centering
  \includegraphics[width=\columnwidth]{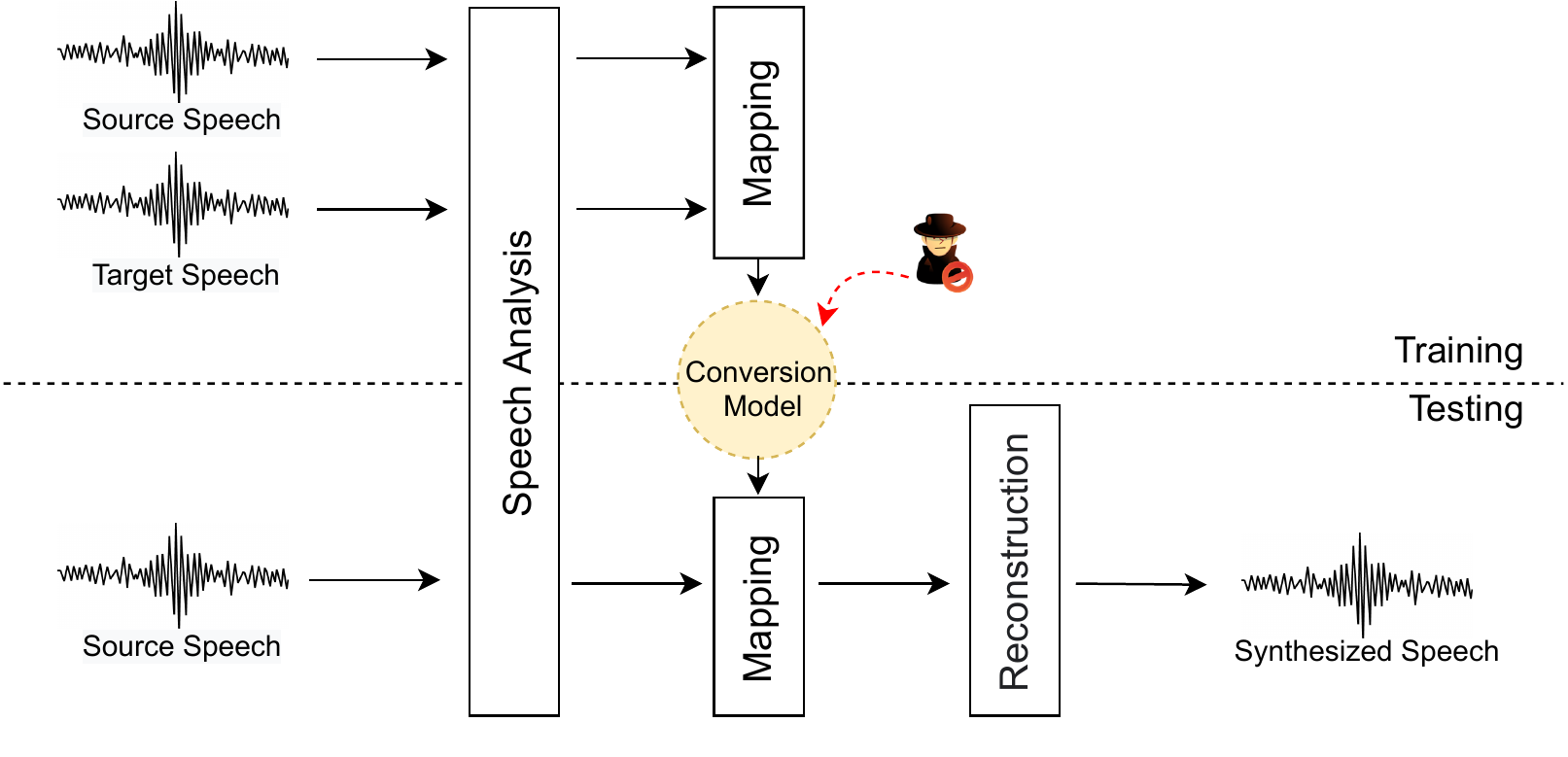}
  \caption{Voice conversion pipeline: (1) for training, the speech signals from the source and target decompose into features, and then feature mapping performs the modification of these features from source to target speaker resulting in conversion model, (2) for testing, the output of the conversion model is used as a vocoder's input to regenerate the speech with the target speaker.}
      \label{fig:E2E}
\end{figure}

%% file: background.tex
\section{Voice Transformation and Authentication}
\label{sec:background}

\subsection{Voice Conversion}
Voice conversion involves multiple speech processing techniques, such as speech analysis, spectral conversion, prosody conversion, speaker characterization, and vocoding. A typical voice conversion pipeline includes speech analysis, mapping, and reconstruction modules. Deep learning techniques also transform the way we implement the analysis-mapping-reconstruction pipeline. The concept of embedding in deep learning provides a new way of deriving the intermediate representation, for example, latent code for linguistic content, and speaker embedding for speaker identity. It also makes the disentanglement of speaker from speech content much easier.

\subsubsection{Speech Analysis.}
From the perspective of speech perception, speaker individuality is characterized at three different levels: segmental, supra-segmental, and linguistic information. The segmental information relates to the short-term feature representations, such as spectrum and instantaneous fundamental frequency (F0). The supra-segmental information describes prosodic features such as duration, tone, stress, rhythm over longer stretches of speech than phonetic units. It is more related to the signal but spanning a longer time than the segmental information. The linguistic information is encoded and expressed through lexical content. 

Voice conversion technology is to deal with the segmental and suprasegmental information while keeping the language content unchanged~\cite{8844600}. The speech analyzer decomposes the speech signals of a source speaker into features that represent supra-segmental and segmental information.

\subsubsection{Mapping.}
The mapping module has taken centre stage in many studies. These techniques can be categorized in different ways, for example, based on the use of training data-parallel vs non-parallel, the type of statistical modeling technique (parametric vs non-parametric), the scope of optimization (frame-level vs utterance level) and the workflow of conversion (direct mapping vs inter-lingual). 

The simplest form of voice conversion (\ie~mapping) requires parallel data for training and it is capable of one-to-one speaker conversion. Parallel data include the same transcription utterances spoken by the source and target speakers and they are highly expensive to collect. Thus, several studies attempted to use non-parallel data to train voice conversion models. In the case of multiple-speaker voice conversion, one-to-one speaker conversion algorithms may be applied to obtain separately trained models for all possible combinations of speaker pairs. However, this approach becomes impractical as the number of speakers increases. Traditional VC research includes modeling spectral mapping with statistical methods such as Gaussian mixture model (GMM), partial least squares regression, and sparse representation. Recent deep learning approaches such as deep neural network (DNN), recurrent neural network (RNN) and generative adversarial network (GAN) have advanced the state-of-the-art. The mapping module changes them towards the target speaker.

\subsubsection{Vocoding.}
Speech reconstruction can be seen as an inverse function of speech analysis that operates on the modified parameters and generates an audible speech signal. It works with speech analysis in tandem. A~\emph{vocoder} learns to reconstruct audio waveforms from acoustic features~\cite{oord2016wavenet}. Traditionally, the waveform can be vocoded from these acoustic or linguistic features using handcrafted models such as WORLD~\cite{morise2016world}, Straight~\cite{kawahara2006straight}, and Griffin-Lim~\cite{griffin1984signal}. However, the quality of those traditional vocoders was limited by the difficulty in accurately estimating the acoustic features from the speech signal. Neural vocoders such as Wavenet~\cite{oord2016wavenet} have rapidly become the most commonly used vocoding method for speech synthesis. Although it improved the quality of generated speech, it has significant cost in computation power and data sources, and suffers from poor generalization~\cite{lorenzo2018towards}. To solve this problem, many architectures such as Wave Recurrent Neural Networks (WaveRNN)~\cite{kalchbrenner2018efficient} have been proposed. WaveRNN combines linear prediction with recurrent neural networks to synthesize neural audio much faster than other neural synthesizers. 

A vocoder is used to express a speech frame with a set of controllable parameters that can be converted back into a speech waveform. Voice conversion systems only modify the speaker-dependent characteristics of speech, such as fundamental frequency (F0), intonation, intensity, and duration, while carrying over the speaker-independent speech content.  The reconstruction module re-synthesizes time-domain speech signals.

\subsection{Speaker Verification Techniques}
Speaker verification is integral to many security applications. This is to verify the identity of a person from the characteristics of the voice. Contemporary ASV systems involve two processes: offline training (\ie~registration or enrollment) and runtime verification. During the offline training, the ASV system uses speech samples provided by the target speaker to extract certain spectral, prosodic, or other high-level features to create a speaker model. Then, in the runtime verification phase, the receiving voice is verified against the trained speaker model~\cite{wang2020asvspoof} and the verification score is compared with a pre-defined threshold. If the score is higher than the threshold, the test is accepted, or rejected otherwise. It is a binary decision task and a verification score is estimated based on the claimed speaker’s model. 

\subsubsection{Speech Analysis.}
Typically, an encoder network extracts frame-level representations from acoustic features (\eg~Mel Frequency Cepstrum Coefficients (MFCCs), filter-banks, or spectrogram). This is followed by a global temporal pooling layer that aggregates the frame-level representation into a single vector per utterance. Finally, a feed-forward classification network processes this single vector to calculate speaker class posteriors~\cite{nagrani2017voxceleb}. Typically, in the evaluation phase, the speaker embedding is extracted from the first affine transform after the pooling layer. Different x-vector systems are characterized by different encoder architectures, pooling methods, and training objectives (\eg~softmax, angular softmax, contrastive, and triplet losses)~\cite{villalba2020state}.\\
\textbf{Traditional Methods.}
Speaker identification was dominated by Gaussian Mixture Models (GMMs) trained on low dimensional feature vectors~\cite{reynolds2000speaker}. The state-of-the-art involves both the use of joint factor analysis (JFA) based methods which model speaker and channel subspaces separately and i-vectors that attempt to model both subspaces into a single compact, low-dimensional space~\cite{dehak2010front}. These systems rely, however, on a low dimensional representation of the audio input, e.g., MFCCs, and thus rapidly degrade in verification performance with real-world noise, and may be lacking in speaker-discriminating features (\eg~pitch information)~\cite{nagrani2017voxceleb}.\\
\textbf{Deep Learning Methods.}
DNN based acoustic models were used instead of the GMM in the i-vector framework~\cite{dehak2010front}. Speaker recognition systems based on Convolutional Neural Networks (CNNs) are often built with off-the-shelf backbones such as VGG-Net or ResNet. An alternative approach is to use DNN to extract bottleneck features~\cite{fu2014tandem, liu2015deep, tian2015investigation} or speaker representations directly~\cite{chen2015multi}. For example, speaker representations such as d-vector~\cite{variani2014deep, heigold2016end} and RNN/LSTM based sequence-vector (s-vector)~\cite{bhattacharya2016deep} have been applied as robust speaker embeddings.

\subsubsection{Speaker Modeling.}
There are two kinds of speaker verification (SV) systems: Text-independent (TI)-SV and Text-dependent (TD)-SV systems. TD-SV assumes cooperative speakers and requires the speaker to speak fixed or spontaneously prompted utterances, whereas TI-SV allows the speaker to speak freely during both enrolment and verification. Both TI-SV and TD-SV systems share the feature extraction techniques while being different in the speaker modeling. However, the text-prompted speaker recognition systems have been the preferred alternative in many practical applications.\\
\textbf{TI-SV Modeling.}
In the text-independent (TI) mode, there are no constraints on the text. Thus, the enrollment and test utterances may have completely different texts. For such cases, it is more convenient for the users to operate. Text-independent ASV systems are more flexible and are able to accept arbitrary utterances, e.g., different languages, from speakers.\\
\textbf{TD-SV Modeling.}
In the text-dependent (TD) mode, the user is expected to speak a pre-determined text for both training and test. Due to the prior knowledge (lexical content) of the spoken phrase, TD systems are generally more robust and can achieve good performance. Text-dependent ASV is more widely selected for authentication applications, since it provides higher recognition accuracy with fewer required utterances for verification.

%% file: taxonomy.tex
\section{Voice Disguise vs. Speaker Authentication Systems}
\label{sec:taxonomy}
This section presents some insights into different types of spoofing attacks, followed by two distinct measurement estimators: Disguise and Anonymization. We then use a tandem framework combining these two estimators to manage the application of privacy-preserving solutions.

\subsection{Generic Attack Model}
\begin{figure}[t!]
  \centering
  \includegraphics[width=\columnwidth]{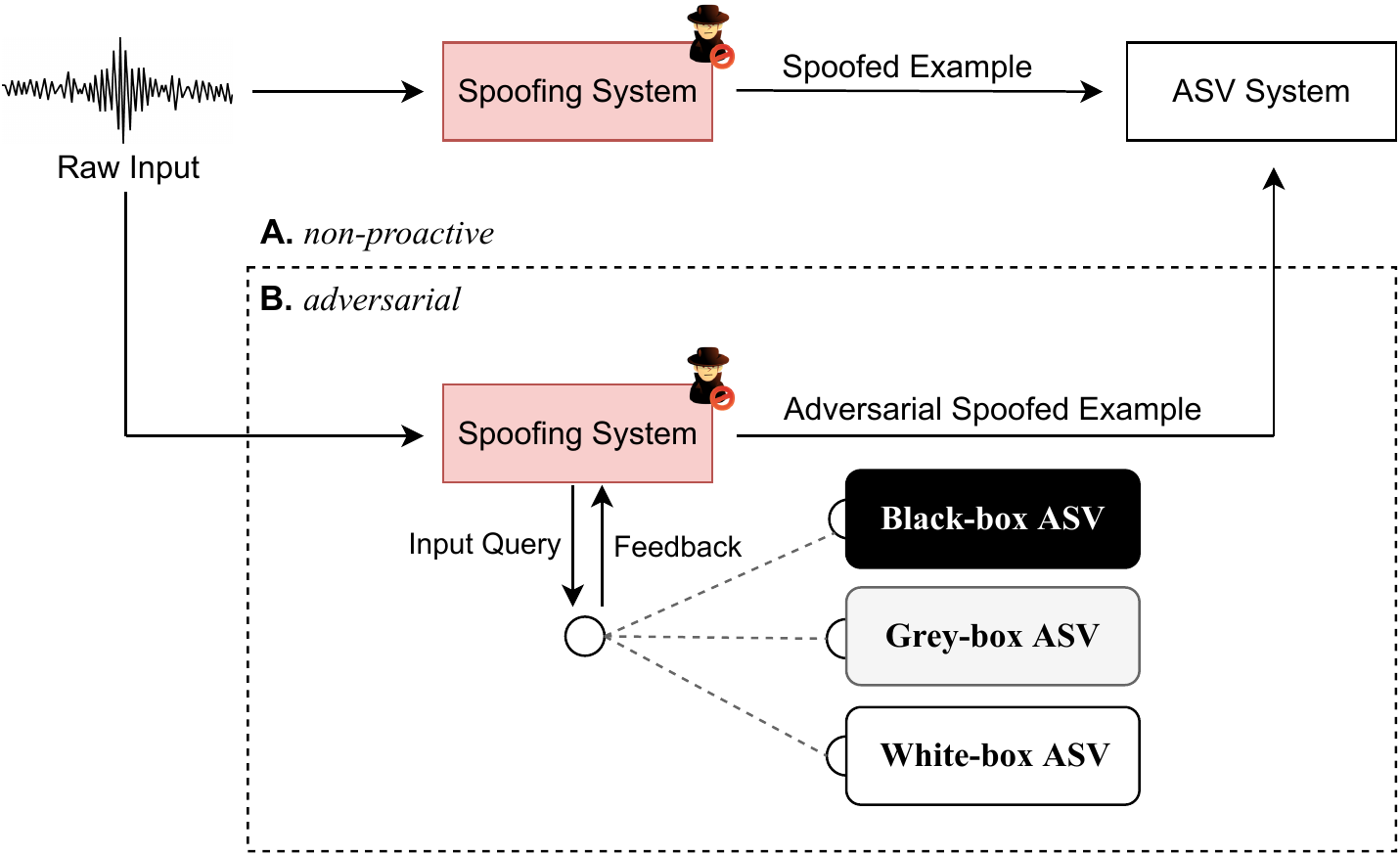}
  \caption{Spoofing attacks (A) non-proactive attacks (B) adversarial attacks: using black-box, grey-box and white-box ASV~\cite{das2020attacker}.}
      \label{fig:spoofing_attacks}
\end{figure}

Security of automatic speaker verification (ASV) systems can be compromised by various spoofing attacks (\eg~speech synthesis and voice conversion). We consider a scenario in which a user seeks to compromise a system or service protected by ASV. It is assumed in these scenarios that the microphone is not controlled by the authentication system and is instead chosen by the user (\ie~post-sensor scenario). An example is that voice spoofing attacks can be used to impersonate a person's voice for voice assistants like Amazon Alexa or Google Assistant to shop online, send messages, control smart home appliances, and grant undesirable access to personal users' data such as financial information. Such attacks are not necessarily fraudulent, whereby a user of these services may want to conceal their identity to spoof or trick third parties for privacy preservation purposes. Attacks then take the form of synthetic speech or converted voice, which is presented to the ASV system without acoustic propagation or microphone effects (\ie~logical-access voice spoofing techniques). From the attacker's perspective, spoofing attacks (\ie~ in our case, assuming the anonymization system operates as a spoofing system) can be categorized into non-proactive and adversarial attacks causing potential threats on ASV, spoofing countermeasures, or both~\cite{das2020attacker}, as shown in Figure~\ref{fig:spoofing_attacks}.

\subsubsection{Spoofing with Non-proactive Attacks}
The attacker lacks a direct optimization target related to the attacked ASV system, as shown in Figure~\ref{fig:spoofing_attacks} (A). Basically, crafting non-proactive attacks represent ideas or technology originally designed for completely different aims and purposes instead of fooling ASV systems~\cite{das2020attacker}. One example is VC and TTS attacks, which aim at modifying source speaker identity to that of a target speaker, and to produce text in a given target speaker’s voice, respectively. VC and TTS technology takes place as a key concept behind a privacy protection mechanism that anonymizes the speaker identity~\cite{tomashenko2020introducing}. Thus, TTS and VC attacks can compromise the security of ASV systems as a side-purpose rather than its original objective in helping, for example, to give those with conditions like autism the ability to speak naturally~\cite{das2020attacker}. In this paper we focus on the non-proactive type, and consider anonymization objective in fooling ASV as a side effect of voice transformation technology.
\subsubsection{Spoofing with Adversarial Attacks}
The attacker leverages the information of the attacked ASV system to generate spoofed samples and can use the knowledge of either the attacked ASV or another similar ASV to generate adversarial samples~\cite{das2020attacker, villalba2020x}, see Figure~\ref{fig:spoofing_attacks} (B). Adversarial attacks can be broadly divided into black, grey, and white-box attacks~\cite{goodfellow2014explaining}. In the black-box setting, the adversary’s observation is limited to the system output (\eg~speaker similarity score) and the model parameters and the intermediate steps of the computation are not accessible to the attacker~\cite{8835245}. In the grey-box, the attacker has some information such as features of the speakers and their implementation, but not their statistical models~\cite{das2020attacker}. The white-box attacks pose the greatest threat as the attackers have full knowledge of the model under attack including its parameters which are needed for prediction~\cite{8835245, nakamura2019v2s}. We assume that anonymization tools are designed without considering specific knowledge about ASV used for authenticating either target or non-target users.

\subsection{Verification-to-Disguise (V2D) Estimation}

\begin{tcolorbox}
\begin{verbatim}
V2D is a spoofing detector to discriminate genuine
and synthetic speech utterances.
\end{verbatim}
\end{tcolorbox}

\begin{figure}[t!]
  \centering
  \includegraphics[width=\columnwidth]{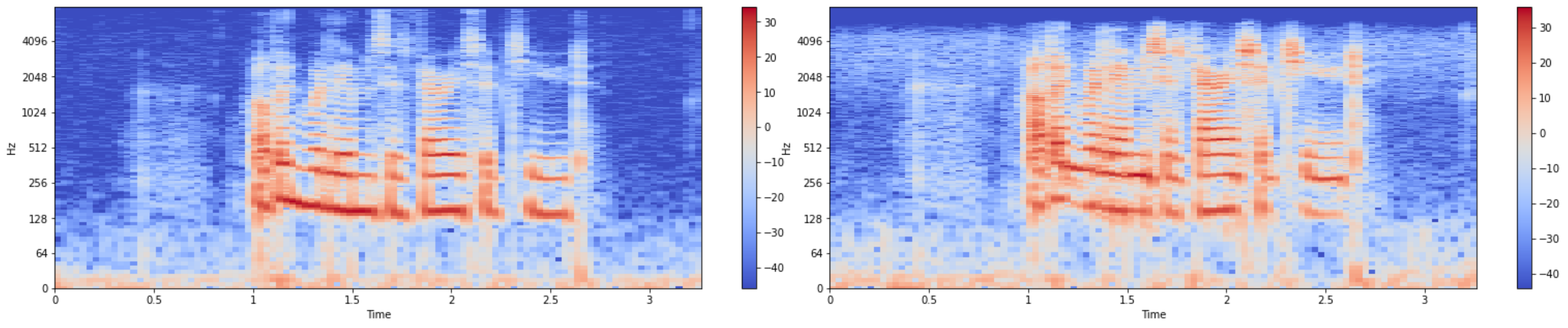}
  \caption{The same text utterance (\emph{"Robin Williams is very subdued." })’s spectral envelope (log scale) of genuine speech contains natural transition (raw, left) while the spoofed speech does not (anonymized, right).}
      \label{fig:spoofing_specto}
\end{figure}

Spoofing countermeasures (CM) are introduced to the ASV systems to protect them from various attacks~\cite{wu2020defense, liu2019adversarial}. High-performance anti-spoofing is used to protect ASV by identifying and filtering spoofing audio that is deliberately generated by text-to-speech, voice conversion, audio replay, etc. Claimed identities are thus only accepted if a test signal attains a countermeasure score lower than its threshold. Therefore, the existing spoofing countermeasure involves the extraction of various parameters of prediction error, aiming to capture the features that will help to differentiate genuine from spoof speech signals. Spoofing countermeasures use particular features that capture the unique aspects of human speech production, under the hypothesis that machines cannot emulate many of the fine-level intricacies of the human speech production mechanism~\cite{9053086}. This could be because of the complexity of the human speech production mechanism, human speech has a greater degree of inconsistency than machine-generated speech, as shown in Figure~\ref{fig:spoofing_specto}. Typically, for deep-learning based voice spoofing detection models, the speech features (\eg~LFCC or CQCC) are fed into a neural network to calculate an embedding vector for the input utterance. The objective of training this model is to learn an embedding space in which the genuine voices and spoofing voices can be well discriminated. The embedding would be further used for scoring the confidence of whether the utterance belongs to genuine speech or not.

By increasing the performance of spoofing countermeasures in detecting disguised voices, the privacy-based transformation methods may not be sufficient to further protect the users' privacy. In our evaluation framework, we evaluate the artifact in the transformed voices by these systems using a spoofing countermeasure to indicate the level of artifacts left in the converted speech. In our case, CM tries to detect whether the privacy-transformed utterance will be detected as spoofed or not. Thus, we may have two initial scenarios regarding the privacy protection level offered by the anonymization solution:~\emph{strong security}, indicating that the transformed input detected as spoofed and will be prevented from accessing the ASV system, and~\emph{weak security}, indicating that the transformed input bypassed the spoofing countermeasures and thus might present security issues to the authentication system. This V2D output will be used in a security estimation over the inputs of VUIs systems.

\subsection{Verification-to-Anonymization (V2A) Estimation}
\begin{tcolorbox}
\begin{verbatim}
V2A is a speaker detector to verify whether the 
given utterance is from the target speaker or not.
\end{verbatim}
\end{tcolorbox}

VUIs use spoken commands to carry out various actions. Sometimes it is necessary to collect some speech to improve and adapt the assistant’s models to the user’s speech. In this case, an attacker could have access to sensitive user data (e.g., they may observe or infer personal information such as identity, age, and gender that can be easily obtained from these utterances). Thus, the objective is privacy preservation, suppressing critical speaker information from speech. To protect their privacy, users may implement a privacy-preservation tool over their data to minimize the personal information, while allowing one or more downstream goals to be achieved. Recent attempts have focused on speech transformation, voice conversion, and speech synthesis as technologies underpinning these solutions.

Privacy by anonymization has achieved remarkable success in concealing identity to preserve users' privacy~\cite{Hidebehind_2018, 255304, 9102875, srivastava2020design, 8844600}. Although the primary purpose is to protect privacy, this has successfully misled the verification systems~\cite{srivastava2020design}. Speakers want to hide their identity while allowing any desired goal to be potentially achieved. In order to hide his/her identity, benign users pass their utterances through an anonymization system before sharing/publication. The resulting anonymized utterances are called trial utterances. They sound as if they were uttered by another speaker, which we call a pseudo-speaker that may be an artificial voice not corresponding to any real speaker. In our case, ASV tries to detect whether the privacy-transformed utterance is spoken by the target speaker or not. Thus, we may have two initial scenarios without considering if a piece of voice is disguised or not:~\emph{better privacy}, indicating that the transformed input is not linkable to the target-speaker,~\emph{worst-case privacy}, indicating that we can still distinguish the target-speaker of the utterance. This V2A output will be used in privacy estimation over the inputs of VUIs systems.

\subsection{Tandem Framework}
\begin{figure}[t!]
  \centering
  \includegraphics[width=\columnwidth]{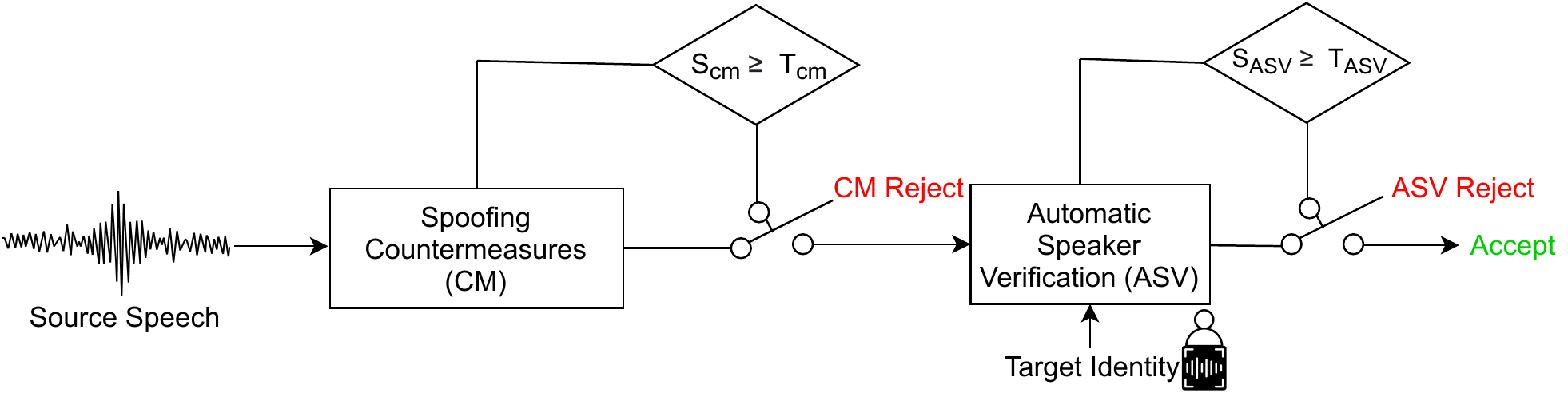}
  \caption{A tandem system consisting of automatic speaker verification (ASV) and spoofing countermeasure (CM) modules. $S_{cm}$, $T_{cm}$ and $S_{asv}$ , $T_{asv}$ denote the scores and thresholds of the CM and ASV systems, respectively.}
      \label{fig:spoofing_score}
\end{figure}

Despite their apparent variations, spoofed inputs and anonymization tools share the same objective of forcing target authentication systems to misclassify pre-defined inputs (target or not). We will focus on the assessment of tandem systems whereby a V2D (\ie~CM) serves as a ‘gate’ to determine whether a given speech input originates from a genuine user, before passing it to V2A (\eg~ASV system). Assuming that verifying the signal integrity comes before achieving privacy, we envision a cascaded (tandem) system where a spoofing countermeasure system is placed before the authentication system (\ie~regarding the anonymization scenario in this paper), to prevent spoofing attacks from reaching this system, as shown in Figure~\ref{fig:spoofing_score}. 

To assess the joint performance of V2D and V2A, we adopt a new metric called (minimum) tandem detection cost function (t-DCF)~\cite{fu2014tandem}. A t-DCF has been proposed by ASVspoof 2019 as its primary performance metric with a focus on the spoofing attack prior. The t-DCF is based on statistical detection theory and involves detailed specification of an envisioned application. It is a parameterized cost that makes the modeling assumptions of an envisioned operating environment (application) explicit. A key feature of t-DCF is the assessment of a tandem system while keeping the two subsystems (CM and authentication) isolated from each other and they can be developed independently of each other. Since the nature of spoofing attacks is never known in advance, t-DCF metric, therefore, reflects the cost of decisions in a Bayes/minimum risk sense by combining a fixed cost model with trial priors. Thus, beyond its practice for spoofing countermeasures, the specification of costs and priors tailors the t-DCF metric towards the development of secure and private applications for a range of different configurations.  

The desired security-privacy trade-off might specify through detection costs assigned to erroneous system decisions and prior probabilities assigned to the commonality of targets, non-targets, and spoofing attacks. For example, a high-security user authentication application (\eg~access control) where target users and spoofing attacks are almost equally likely to occur, while non-target users are rare. False acceptances (\ie~whether of non-targets or VC attacks) incur a ten-fold cost relative to false rejections. The higher the t-DCF value, the more detrimental the spoofing attack. The maximum value of 1.0 indicates an attack that renders the tandem system useless. Thus, this trade-off can have three possible results which are: (1) CM bonafide and ASV accept means `high privacy', (2) CM bonafide and ASV reject, which means `low privacy', and (3) CM spoof and ASV accept/reject. Therefore, we want to confirm whether this objective metric can capture such score variations and predict scores for evaluating VUIs robustness and privacy.

%% file: experiments.tex
\section{experiments}
\label{sec:experiment}

In this section, we describe the datasets, neural network architectures, and corresponding attacks \& countermeasure settings that we use in our experiments. 

\subsection{Study Setting}
\begin{table}[]
\caption{Details of the used VC systems aiming to anonymize the speaker identity}
\label{tab:anon}
\small
\begin{tabular}{c|ll}
\hline
System & VC model & Vocoder \\ \hline
P1 & VoicePrivacy Challenge & Neural Source-filter \\
P2 & VQVAE & World\\
P3 & VQVAEGAN & Parallel WaveGAN\\
P4 & CycleVQVAE & ParallelWaveGAN \\
P5 & CycleVQVAEGAN & Parallel WaveGAN \\ \hline
\end{tabular}
\end{table}
\textbf{Datasets.} To factor out the influence of specific datasets, we primarily use 4 benchmark datasets:\\
\textbf{VoxCeleb.}
VoxCeleb dataset~\cite{nagrani2017voxceleb} contains over 100,000 utterances for 7325 celebrities, extracted from videos uploaded to YouTube. The speakers span a wide range of different ethnicities, accents, professions and ages. It was curated to facilitate the development of automatic speaker recognition systems. We use it to train and evaluate the authentication system.\\
\textbf{VCTK.}
VCTK dataset~\cite{yamagishi2019vctk} includes speech data uttered by 110 English speakers with various accents. Each speaker reads out about 400 sentences. It was recorded for the purpose of building HMM-based text-to-speech synthesis systems, especially for speaker-adaptive HMM-based speech synthesis using average voice models trained on multiple speakers and speaker adaptation technologies. We use it to train and evaluate the anonymization systems.\\
\textbf{VCC2020.}
VCC2020 dataset~\cite{Yi2020} is based on the Effective Multilingual Interaction in Mobile Environments (EMIME) dataset~\cite{wester2010emime}, which is a bilingual database of Finnish/English, German/English, and Mandarin/English data. There are seven male and seven female speakers for each language, English, Finnish, German, and Mandarin, ending up in 56 speakers in total. It uses to train and evaluate the anonymization systems.\\
\textbf{ASVspoof2019.}
ASVspoof2019 database~\cite{wang2020asvspoof} for logical access is based upon a standard multi-speaker speech synthesis database called VCTK~\cite{yamagishi2019vctk}. Genuine speech is collected from 107 speakers (46 male, 61 female) and with no significant channel or background noise effects. Spoofed speech is generated from the genuine data using a number of different spoofing algorithms. It uses to train and evaluate the spoofing countermeasure systems.
\newline
\newline
\textbf{Anonymization.}
As VC is a basic technique behind most current state-of-the-art anonymization solutions (\ie~offering identity privacy preservation)~\cite{tomashenko2020introducing, Hidebehind_2018, 255304, 9102875, champion2021study, srivastava2020design, 9247219, aloufi2019emotionless}, we implement the following five systems (\ie~P1-P5 respectively): `VoicePrivacy Baseline', `VQVAE', `VQVAEGAN', `CycleVQVAE', and `CycleVQVAEGAN'. For `P1', we use the baseline implementation for the VoicePrivacy challange. Then, we use an open-source nonparallel VC software named crank~\cite{kobayashi2021crank} to implement various VC systems with different configurations including hierarchical architectures `P2', generative adversarial networks `P3', cyclic architectures `P4', speaker adversarial training `P5', and neural vocoders, as shown in Table~\ref{tab:anon}. Following a typical VC systems pipeline in these systems, several steps such as preparing the dataset, feature extraction, training, and conversion are implemented in order to reconstruct the speech utterance while transforming the speaker identity.
\newline
\newline
\textbf{Authentication System.}
We use an x-vector~\cite{snyder2018x} embedding extractor network that was a pre-trained recipe of the Kaldi toolkit~\cite{povey2011kaldi}. Training was performed using the speech data collected from 7325 speakers contained in the entire VoxCeleb2 corpus~\cite{nagrani2017voxceleb}. We extract 512-dimensional x-vectors which are fed to a probabilistic linear discriminant analysis (PLDA). PLDA scoring is used to make a rejection/acceptance decision about the speaker identity.
\begin{table}[]
\caption{Categorical tags of worst-case privacy disclosure~\cite{Nautsch_2020} based on the decision made by an adversary, the better an adversary can make decisions, despite the privacy preservation is applied, the worse is the categorical tag.}
\label{tab:zebra}
\begin{tabularx}{\columnwidth}{cll}
\hline
Tag & \multicolumn{1}{c}{Category}                  & \multicolumn{1}{c}{Posterior odds ratio (flat prior)}   \\ \hline
0   & \multicolumn{1}{c}{\textbf{$l = 1 = 10^{0}$}} & 50:50 decision making of the adversary                  \\ \hline
A   & $10^{0} \leq l < 10^{1}$                      & adversary better decisions than 50:50             \\
B   & $10^{1} \leq l < 10^{2}$                      & one wrong decision in 10 to 100       \\
C   & $10^{2} \leq l < 10^{4}$                      & one wrong decision in 100 to 1000     \\
D   & $10^{4} \leq l < 10^{5}$                      & one wrong decision in 1000 to 100.000 \\
E & $10^{5} \leq l < 10^{6}$ & one wrong decision in 100.000 to 1.000.000 \\
F   & $10^{6} \leq l$                               & one wrong decision in at least 1.000.000  \\ \hline
\end{tabularx}
\end{table}
\newline
\newline
\textbf{Spoofing Countermeasures.} 
Following~\cite{wang2020asvspoof}, we use several countermeasure models based on the light convolutional neural network (LCNN)~\cite{wu2020light}. These models trained on the ASVspoof 2019 logical access scenario considering current strategies that deal with input trials of varied length. We consider three network structures: `LCNN-trim-pad', `LCNN-attention', and `LCNN-lstm-sum'. The loss function can be either AM-softmax, OC-softmax, sigmoid, or MSE for P2SGrad. For more details, refer to~\cite{wang2020asvspoof}. We compare their performance on the ASVspoof2019 logical access (LA) dataset (\ie~ as a known attack (AK)~ \eg~waveform filtering, griffinlim, and spectral filtering) and the anonymized recordings (\ie~as an unknown attack implementing different conversion and vocoder models from those used in the training) across the front ends based on linear frequency cepstral coefficients (LFCCs), linear filter bank coefficients (LFBs), and spectrograms. The LFCC is 60-dimensional extracted from a frame length of 20~ms, a frame shift of 10~ms, a 512-point FFT, a linearly spaced triangle filter bank of 20 channels, and delta plus delta-delta coefficients (\ie~the first dimension replaced by log spectral energy). The LFB has a similar configuration but contains only static coefficients from 60 linear filter-bank channels. The spectrogram configures similarly and has 257 dimensions.
\newline
\newline
\textbf{Measures.} State-of-the-art CM and ASV methods are subsequently utilized to objectively evaluate the impact of voice disguise (\ie~spoofing efficacy) and anonymization level (\ie~privacy protection) by equal error rates (EER) and tandem detection cost function (t-DCF)~\cite{kinnunen2020tandem}.\\
\emph{Spoofing Efficacy.} We measure the attack efficacy by the decision score confidence, which is the probability that the spoofed input belongs to the genuine class as predicted by CMs. We consider the attack successful if the decision score confidence exceeds a threshold.\\
\emph{Privacy Protection.} We measure the level of the privacy protection offered by the anonymization solutions by the decision score confidence, which is the probability that the speech input belongs to the target speaker as predicted by ASV. We estimate the protection by anonymization in terms of the average protection afforded to a population and a worst-case to an individual.

\begin{figure}[t!]
  \centering
  \includegraphics[scale=0.45]{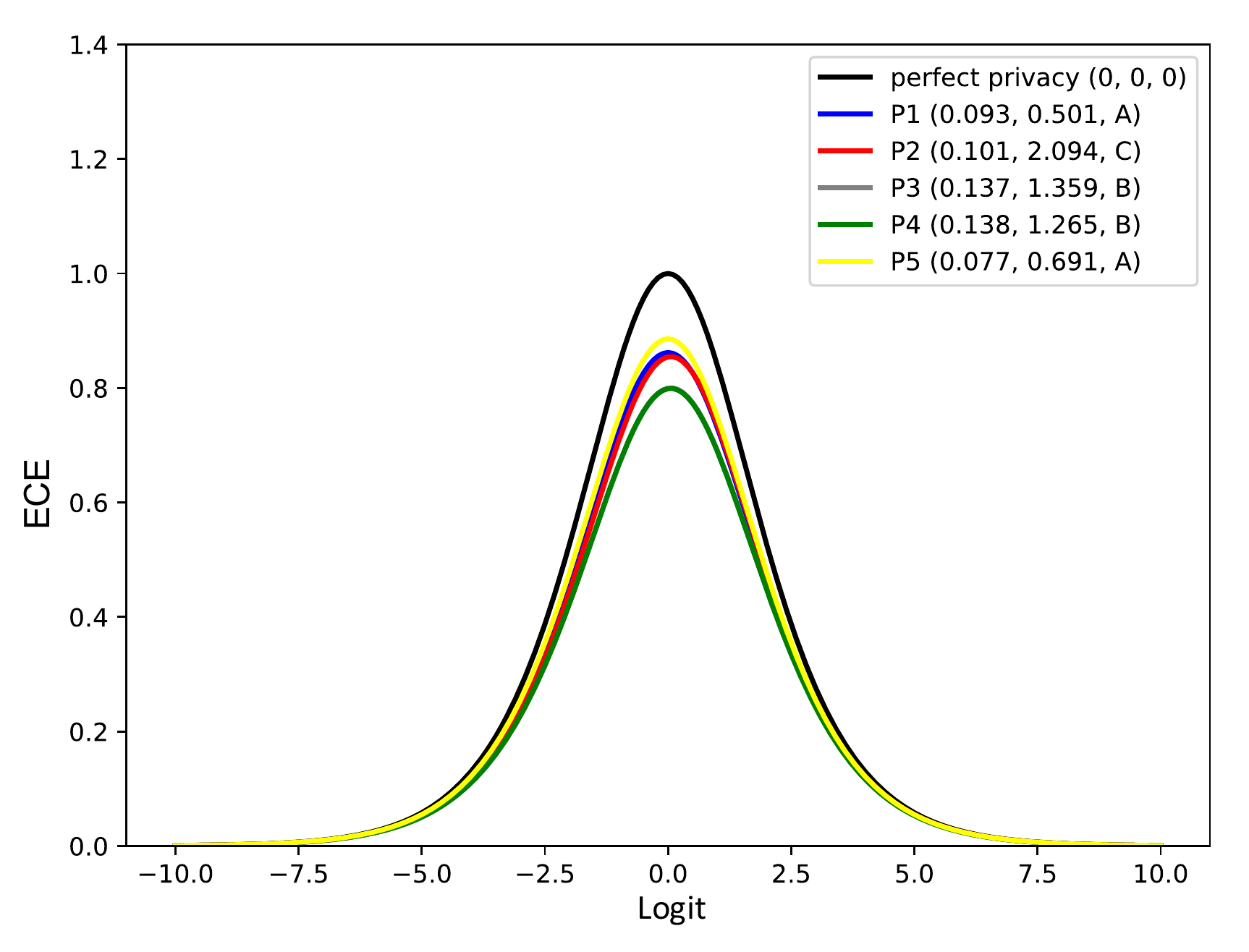}
  \caption{Privacy analysis of the adopted anonymization systems using ZEBRA profile metrics in form of: system name followed by population, individual, and tag values.}
      \label{fig:privacy_protection}
\end{figure}

\subsection{Performance Analysis}

\begin{table*}[]
\small
\caption{EERs on the generated voices by the transformation systems across various CMs applied various features, NN architectures, and loss functions (\ie~ lower EERs is better).}
\label{tab:eer}
\begin{tabularx}{\textwidth}{cl|cccc|cccc|cccc|cccc}
\hline
\multicolumn{1}{l}{} &
   &
  \multicolumn{4}{c|}{AM-softmax} &
  \multicolumn{4}{c|}{OC-softmax} &
  \multicolumn{4}{c|}{Sigmoid} &
  \multicolumn{4}{c}{P2SGrad} \\ \hline
\multicolumn{1}{c|}{Feature} &
  \multicolumn{1}{c|}{NN} &
  KA &
  P1 &
  P2 &
  P3-P5 &
  KA &
  P1 &
  P2 &
  P3-P5 &
  KA &
  P1 &
  P2 &
  P3-P5 &
  KA &
  P1 &
  P2 &
  P3-P5 \\ \hline
\multicolumn{1}{c|}{\multirow{3}{*}{LFB}} &
  L-T-P &
  5.580 &
  0.410 &
  0.320 &
  0.330 &
  5.980 &
  2.630 &
  0.350 &
  0.360 &
  7.000 &
  2.020 &
  \textbf{0.160} &
  \textbf{0.160} &
  6.810 &
  2.000 &
  0.220 &
  0.230 \\
\multicolumn{1}{c|}{} &
  L-A &
  4.250 &
  2.110 &
  \textbf{0.210} &
  0.230 &
  4.010 &
  1.210 &
  \textbf{0.170} &
  0.190 &
  3.340 &
  1.190 &
  0.270 &
  0.190 &
  3.980 &
  2.340 &
  \textbf{0.130} &
  \textbf{0.130} \\
\multicolumn{1}{c|}{} &
  L-L-S &
  4.230 &
  3.170 &
  0.220 &
  0.230 &
  5.810 &
  3.640 &
  0.230 &
  0.230 &
  7.040 &
  3.210 &
  0.260 &
  0.260 &
  5.060 &
  1.060 &
  0.290 &
  0.290 \\ \hline
\multicolumn{1}{c|}{\multirow{3}{*}{SPEC}} &
  L-T-P &
  4.840 &
  53.00 &
  32.44 &
  37.49 &
  4.410 &
  33.01 &
  24.48 &
  24.78 &
  3.090 &
  21.00&
  12.50 &
  12.50 &
  2.940 &
  1.150 &
  \textbf{0.420} &
  \textbf{0.420} \\
\multicolumn{1}{c|}{} &
  L-A &
  4.020 &
  6.230 &
  \textbf{4.600} &
  5.730 &
  4.050 &
  12.08 &
  4.780 &
  5.080 &
  3.920 &
  6.310 &
  6.250 &
  6.250 &
  4.720 &
  6.040 &
  5.270 &
  6.020 \\
\multicolumn{1}{c|}{} &
  L-L-S &
  3.960 &
  14.49 &
  6.200 &
  11.07 &
  2.810 &
  1.000 &
  \textbf{0.420} &
  0.450 &
  3.290 &
  3.000 &
  1.040 &
  \textbf{0.910} &
  2.370 &
  1.490 &
  0.780 &
  1.460 \\ \hline
\multicolumn{1}{c|}{\multirow{3}{*}{LFCC}} &
  L-T-P &
  3.040 &
  27.06 &
  18.74 &
  18.76 &
  2.930 &
  9.320 &
  5.530 &
  5.920 &
  2.500 &
  7.120 &
  6.250 &
  6.250 &
  2.310 &
  6.170 &
  6.020 &
  \textbf{5.660} \\
\multicolumn{1}{c|}{} &
  L-A &
  2.990 &
  8.210 &
  6.250 &
  7.110 &
  2.910 &
  7.070 &
  6.250 &
  6.250 &
  3.180 &
  6.030 &
  \textbf{5.790} &
  5.960 &
  2.720 &
  6.320 &
  6.250 &
  5.890 \\
\multicolumn{1}{c|}{} &
  L-L-S &
  2.460 &
  6.010 &
  \textbf{5.240} &
  5.370 &
  2.230 &
  6.110 &
  \textbf{5.370} &
  5.610 &
  2.670 &
  6.470 &
  7.420 &
  6.250 &
  1.920 &
  6.340 &
  6.250 &
  6.250 \\ \hline
\end{tabularx}
\end{table*}

\begin{table*}[]
\small
\caption{min t-DCFs (\ie~joint performance with ASV) on the generated voices by the transformation systems across various CMs applied various features, NN architectures, and loss functions.}
\label{tab:t-DCF}
\begin{tabularx}{\textwidth}{cl|cccc|cccc|cccc|cccc}
\hline
\multicolumn{1}{l}{} &
   &
  \multicolumn{4}{c|}{AM-softmax} &
  \multicolumn{4}{c|}{OC-softmax} &
  \multicolumn{4}{c|}{Sigmoid} &
  \multicolumn{4}{c}{P2SGrad} \\ \hline
\multicolumn{1}{c|}{Feature} &
  \multicolumn{1}{c|}{NN} &
  KA &
  P1 &
  P2 &
  P3-P5 &
  KA &
  P1 &
  P2 &
  P3-P5 &
  KA &
  P1 &
  P2 &
  P3-P5 &
  KA &
  P1 &
  P2 &
  P3-P5 \\ \hline
\multicolumn{1}{c|}{\multirow{3}{*}{LFB}} &
  L-T-P &
  0.120 &
  0.007 &
  0.006 &
  0.007 &
  0.150 &
  0.007 &
  0.007 &
  0.007 &
  0.160 &
  0.013 &
  0.006 &
  0.013 &
  0.170 &
  0.015 &
  0.006 &
  0.015 \\
\multicolumn{1}{c|}{} &
  L-A &
  0.110 &
  0.006 &
  0.006 &
  0.006 &
  0.110 &
  0.020 &
  0.023 &
  0.013 &
  0.060 &
  0.014 &
  0.012 &
  0.014 &
  0.080 &
  0.014 &
  0.013 &
  0.013 \\
\multicolumn{1}{c|}{} &
  L-L-S &
  0.090 &
  0.015 &
  0.015 &
  0.015 &
  0.160 &
  0.018 &
  0.013 &
  0.015 &
  0.180 &
  0.015 &
  0.005 &
  0.005 &
  0.140 &
  0.006 &
  0.006 &
  0.006 \\ \hline
\multicolumn{1}{c|}{\multirow{3}{*}{SPEC}} &
  L-T-P &
  0.120 &
  0.492 &
  0.440 &
  0.510 &
  0.123 &
  0.680 &
  0.580 &
  0.670 &
  0.085 &
  0.184 &
  0.160 &
  0.210 &
  0.085 &
  0.016 &
  0.008 &
  0.008 \\
\multicolumn{1}{c|}{} &
  L-A &
  0.110 &
  0.054 &
  0.031 &
  0.052 &
  0.105 &
  0.058 &
  0.030 &
  0.039 &
  0.109 &
  0.240 &
  0.230 &
  0.190 &
  0.135 &
  0.058 &
  0.040 &
  0.058 \\
\multicolumn{1}{c|}{} &
  L-L-S &
  0.101 &
  0.120 &
  0.069 &
  0.113 &
  0.077 &
  0.018 &
  0.008 &
  0.009 &
  0.087 &
  0.037 &
  0.020 &
  0.018 &
  0.060 &
  0.031 &
  0.015 &
  0.029 \\ \hline
\multicolumn{1}{c|}{\multirow{3}{*}{LFCC}} &
  L-T-P &
  0.068 &
  0.290 &
  0.260 &
  0.260 &
  0.068 &
  0.059 &
  0.048 &
  0.056 &
  0.069 &
  0.218 &
  0.191 &
  0.188 &
  0.056 &
  0.080 &
  0.063 &
  0.050 \\
\multicolumn{1}{c|}{} &
  L-A &
  0.074 &
  0.274 &
  0.220 &
  0.191 &
  0.066 &
  0.122 &
  0.087 &
  0.081 &
  0.068 &
  0.080 &
  0.053 &
  0.057 &
  0.079 &
  0.116 &
  0.073 &
  0.055 \\
\multicolumn{1}{c|}{} &
  L-L-S &
  0.057 &
  0.059 &
  0.042 &
  0.044 &
  0.064 &
  0.051 &
  0.044 &
  0.049 &
  0.064 &
  0.180 &
  0.127 &
  0.110 &
  0.052 &
  0.169 &
  0.144 &
  0.150 \\ \hline
\end{tabularx}
\end{table*}

\subsubsection{V2D Performance}
To evaluate the performance of the spoofing countermeasure system, we use the countermeasure decision score which indicates the similarity of the given utterance with genuine speech. equal error rates (EER) is calculated by setting a threshold on the countermeasure decision score, such that the false alarm rate is equal to the miss rate. A high EER indicates the converted speech to be more human-like speech, whereas a lower EER is the better spoofing countermeasure system at detecting spoofing attacks (\ie~EER is constrained between 0 and 0.5, and values larger than 0.5 indicate decisions worse than random guessing). Then, the t-DCF metric is utilized to assess the influence of CM systems on the reliability of an ASV system. The lower the t-DCF is, the better reliability of ASV is achieved.

In Tables~\ref{tab:eer} and~\ref{tab:t-DCF}, we summarize and compare the approaches to deal with varied-length input and several loss functions reported in the recent speech anti-spoofing literature. We list the EERs and t-DCFs on the evaluation set of ASVspoof19 (LA) and the output of the five conversion systems. Results on loss functions demonstrate the loss function based on sigmoid and P2SGrad have a competitive performance over unknown attacks which achieved a lower equal error rate of 0.16\% and 0.13\% compared to the other loss functions.

\subsubsection{V2A Performance}
A speaker verification system automatically accepts or rejects a claimed identity of a speaker based on a speech sample. Three metrics are estimated: EER and log-likelihood
ratio (LLR) scores, $C_{llr}$~\cite{cll} (\ie~relates to empirical cross-entropy) $C_{llr}$ and $C_{llr}^{min}$. Denoting by $P_{fa}$($\theta$) and $P_{miss}$ ($\theta$) the false alarm and miss rates at threshold $\theta$, the EER corresponds to the threshold $\theta$EER at which the two detection error rates are equal, i.e., EER = $P_{fa}$ ($\theta_{EER}$) = $P_{miss}$ ($\theta_{EER}$). Then, we use the output scores within both: tandem (\ie~EER) and `ZEBRA' frameworks~\cite{Nautsch_2020} (\ie~all scores). The tandem framework serves as a guide for the optimization of a countermeasure for a given authentication system. `ZEBRA' framework measures the average level of privacy protection afforded by a given privacy-preserving solution for a population and the worst-case privacy disclosure for an individual.

\subsubsection{Tandem Framework Performance}
\begin{figure}[t!]
  \centering
  \includegraphics[scale=0.40]{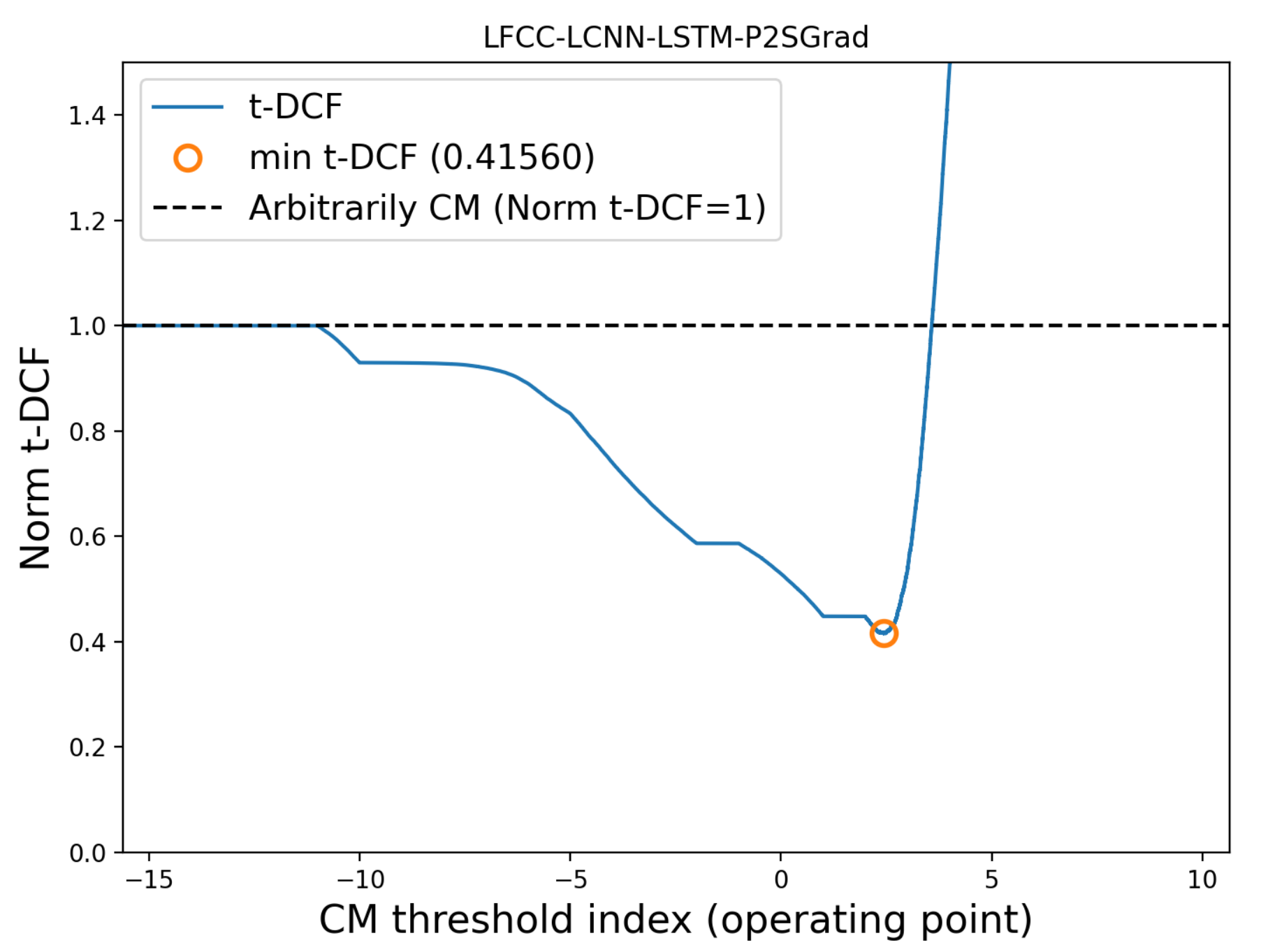}
  \caption{Normalized t-DCF function calculated using: t-DCF parameters and ASV errors, and the threshold of each evaluated CM setting to its optimal value corresponding to perfect calibration over evaluation sets (\ie~t-DCF as a function of the CM threshold). Above, the performance of giving CM~\eg~`LCNN-lstm-sum-P2SGard' compared to bad arbitrary CM.}
      \label{fig:tendam}
\end{figure}

By combining spoofing detection scores with ASV scores, adoption of the t-DCF, we evaluate the impact of spoofing and the performance of spoofing countermeasures upon the reliability of ASV. Both the CM and ASV subsystems will make classification errors. The aim is to assess the performance of the tandem system as a whole taking into account not only the detection errors of both subsystems, but assumed prior frequencies. For `spoofing performance' assessment, we consider the ASV and CM results jointly. Minimum normalized t-DCF, defined as: $t-DCF_{norm}^{min}$ = $t-DCF_{norm} (s_*)$ where $s_* = arg min_s$  $t-DCF_{norm} (s)$ is the optimal threshold determined from the giving evaluation dataset using the ground truth. The t-DCF metric has 6 parameters: (i) false alarm and miss costs for both systems, and (ii) prior probabilities of target and spoof trials (with an implied third, nontarget prior). Specifically, VC audio files may be rejected by a CM system if the audio contains detectable artifacts. Even if VC audio files are passed on to the CM system, they may still be rejected by the ASV if their speaker similarity is not close enough to the target speakers, as shown in Fig.~\ref{fig:tendam}.

\subsubsection{Signal Performance.} Most of the important factors that may clearly affect the performance of speech processing systems, including speaker recognition or even spoofing countermeasure performance are:~\emph{Issues on Background Noise} and~\emph{Issues on Variability}. Background noise is an issue being highlighted by~\cite{zheng2017robustness} as problematic because during training, the speaker often speaks in a clean environment. In contrast, during testing, the speaker speaks in a noisy condition. It disturbs the evaluation test and degrades the performance of the speaker recognition system. Voice variability, also known as session variability, is another factor that may affect the performance of these systems; which can be further classified as intra-variation and inter-variation. Intra-variation occurs due to various factors, such as emotions, rate of utterances, mode of speech, disease, the speaker's mood, and the emphasis given to the word. Inter-variation exists due to anatomical differences in the speech signals due to different transmission channels, such as the different types of microphones and headphones used during the recording of speech utterances, where speech data is sampled at either 8~kHz, 16~kHz, or 22~kHz. Thus, to avoid the models having a mismatched condition, in our experiments we adopt 16~kHz for the used recordings cross all the training and evaluation systems.

\subsection{Implications}

\subsubsection{Effect I: Security Consequences}
To analyze the impact of the privacy-preserving solutions, which rely on anonymity using voice transformation tool, we investigate the following question:~\emph{to what extent can anonymization solutions result in high spoofing risk for ASV and CM?}

\emph{Approch.} Authentication systems are prone to be intentionally fooled using spoofing attacks (\ie~replay, text-to-speech (TTS), and voice conversion (VC)). In our experiments, we involve both spoofed voices synthesized by modern TTS (\ie~using ASVspoof2019 (LA) evaluation set) and VC models. We used converted voices produced by each of the VC systems (\ie~representive of identity anonymization tools).

\emph{Observations.} 
Interestingly, it is noted that despite the difference in the results compared to the known attacks, the countermeasure systems are still able to recognize the converted output (\ie~anonymized) as spoofed inputs. 
Although identity may be pretended by VC, it might be exploited in inappropriate ways (\eg~deepfake problems such as synthesized fake voice). The reason for this may be that all the conversion mechanisms share the need for vocoders to reconstruct waveforms, and even with the development of such techniques, they still can be distinguished compared to the raw utterances.

\emph{Limitations \& Future Work.} 
However, the remaining question is about the robustness of such systems under adversarial attacks,~\emph{are these countermeasures for ASV robust enough to defend against adversarial examples?} For the robustness, we want to point out that two important factors may affect the performance of these systems: (1) adversarial inputs and (2) real-world perturbations (\ie~noisy background). Liu~\cite{liu2019adversarial} start to highlight the vulnerability of some of spoofing countermeasures under both white-box and black-box adversarial attacks with the fast gradient sign (FGSM) and the projected gradient descent (PGD) methods. While Chettri~\etal~in~\cite{chettri2020subband} spot the effect of the real-time perturbations on CMs, which could be due to (1) variations within the spoof class~\eg~speech synthesizers not present in the training set, (2) within the bonafide class~\eg~due to content and speaker, or (3) additional nuisance factors~\eg~background noise. Thus, the ideal CM should generalize across environments, speakers, languages, channels, and attacks to allow maximum utility across different applications. It is an important direction to be explored in the future and we seek to include it 
in our evaluation toward designing secure and private VUIs systems. We seek to use the tandem framework to evaluate a system beyond the ASV (\ie~non-biometric) to add a level of security prior privacy-preserving applications.

\subsubsection{Effect II: Privacy Concerns.}
To analyze the effectiveness of the voice transformation in designing privacy-preserving solutions, we investigate the following question:~\emph{does hiding identity offer an ideal solution for privacy protection in VUI-based services?}

\emph{Approach.}
Due to privacy concerns, oftentimes such data must be de-identified or anonymized before it is used or shared. Therefore, we measured the level of privacy provided by these solutions. Specifically, according to the privacy goal adopted by these solutions, the extent to which the adversary can obtain the identity of the speaker.

\emph{Observations.}
We use the authentication confidence scores cross x-vectors (\ie~V2A) to calculate empirical cross-entropy (ECE). Then, to quantify levels of privacy preservation, the ZEBRA framework uses ECE value to show: (1) the performance on the population level, as the expected ECE is quantified by integrating out all possible prior beliefs (\ie~0 bit) for full privacy, and (2) the performance on the individual level, the worst-case strength of evidence is the maximum(absolute(LLR)). As shown in Figure~\ref{fig:privacy_protection}, ECEs are presented to simulate all possible prior beliefs (\ie~average/expected performance of the presented privacy-preserving solutions). 

\emph{Limitations \& Future Work.} 
Recently, The VoicePrivacy initiative~\cite{tomashenko2020introducing} has promoted the development of anonymization methods that aim to suppress personally identifiable information in speech (\ie~speaker identity), while leaving other attributes such as linguistic content intact. Despite the appeal of anonymization techniques and the urgency to address privacy concerns in the speech domain, the current solutions may be useful from singular perspectives and for achieving a specific goals, like hiding the identity of the speaker, but might fail to sufficiently address other scenarios. Additionally, it would be concerning that if a benign user can achieve privacy by a transformation model, that also entails that a malicious user can break security by bypassing the spoofing countermeasure mechanism. Besides the spoofing perspective, Srivastava~\etal~in~\cite{9053868} found that when the attacker has complete knowledge of the VC scheme and target speaker mapping, none of the existing VC methods will be able to protect the speaker identity, and they classified the attacks based on the attacker’s knowledge about the anonymization method (i.e., ignorant, informed, and semi-informed).

%% file: discussion.tex
\section{Potential Countermeasures}
\label{sec:discussion}
In this section, we highlight potential defenses considering both security and privacy, and then discuss whether the conflicting between them is necessarily or not.

\subsection{Defense}
VUI technologies that allow users to speak to interact with their devices are based on accurate speaker and speech recognition to ensure appropriate responsiveness. While VUIs are offering new levels of convenience and changing the user experience, these technologies raise new and important security and privacy concerns (\eg~spoofing attacks~\cite{13}, false activations~\cite{PETS_2020smarthome}, attributes attack~\cite{aloufi2020privacypreserving}). Thus, in the following, we discuss potential defenses that can reduce the risk of spoofing attacks while maintaining privacy.

\subsubsection{Privacy by Personalization}
In many real-world usage scenarios, our voice may often also contains indicators of our identity, mood, emotions, physical and mental wellbeing that may be used to manipulate us and/or shared with third parties. This raises privacy concerns owing to the capture and processing of voice recordings that may involve two or more people, and without their explicit consent. This violates GDPR provisions, for instance. Furthermore, the deep acoustic models used to analyze these recordings may encode more information than needed for the task of interest (\ie~ASR), such as profiles of users' demographic categories, personal preferences, emotional states, etc., and may therefore significantly compromise their privacy. Current works focus on protecting/anonymizing speaker identity using VC-based mechanisms~\cite{Hidebehind_2018, srivastava2020design}. Based on our results in Table~\ref{tab:eer} and~\ref{tab:t-DCF}, we show the limitation of these techniques in achieving secure VUIs services while maintaining user privacy.

Protecting users’ privacy where speech analysis is concerned continues to be a particularly challenging task. Specifically, privacy-preserving solutions should clearly specify the privacy-utility trade-off in a transparent way: what to protect and what task to achieve. These solutions must be compatible without compromising the security of these systems. This opens a new possibility for on-device personalization of speech processing models, where personalized models are trained on users’ devices. For example, a combination of federated learning and differential privacy has been proposed to develop on-device speaker verification~\cite{paulik2021federated, sim2019investigation}. Adversarial training can be one solution in learning the representation related to the task of interest. Srivastava~\etal~in~\cite{srivastava2019privacy} proposed an on-device encoder to protect the speaker identity using adversarial training to learn representations that perform well in ASR while hiding speaker identity. Likewise, recent applications have suggested the implementation of disentanglement~\cite{aloufi2020privacypreserving} in learning speech representations can enhance the robustness of speech representations and overcome common speaker recognition issues like spoofing attacks~\cite{peri2020empirical}. We hypothesize that learning of speech representation (\ie~task-specific) on devices yields a desirable model that meeds the needs of individual users, and thus can be achieved in a personalised, privacy-preserving way by fine-tuning a global model using standard optimization methods on data stored locally on a single device overcoming the current need of using VC-based mechanism.

\subsubsection{Configurable Privacy} 
Protecting privacy requires more than hiding speaker information or running on-device ASR. Privacy is subjective, with varying attitudes between users, and which may even depend on the services (and/or service providers) with which these systems communicate. Thus, current solutions may be useful from singular perspectives and for achieving a specific goal, like the identity of the speaker, but might fail to sufficiently address configurable privacy. Recently, Aloufi~\etal~in~\cite{aloufi2021configurable} advocate the principle of \textit{configurable privacy}, emphasizing the importance of enabling different privacy settings for optimizing the privacy-utility trade-off and promoting transparent privacy management practices.

Based on our experiments, the output of the tandem framework (\ie~combining verification and authentication) can be helpful in deciding/controlling where to deploy a privacy-preserving solution. For example, assuming such a service does not require authentication (\eg~sharing on social media platforms), then we still need to verify the input speech as a genuine utterance, but a decision threshold of the ASV can be configured to enable access to a non-target speaker. Or, if authentication is required (\eg~smart assistance) the decision threshold can only accept the target speaker (\ie~the results in Figure~\ref{fig:tendam} assume the latter case). In addition, this tandem framework could be also useful if the data owner and the service provider have an agreement on what privacy-preserving/anonymization mechanism to implement it on the shared data, then such a tandem system should enable the input generated by this mechanism while restricting other inputs.

\subsubsection{Online Watermarking}
The concept of speech watermarking (\ie~voice signature) has risen to be an efficient and promising solution to safeguard voice signals. It encrypted a user's personal information into the voice as an inaudible watermark~\cite{9316175}. For example, fingerprinting the audio sample using the acoustics features and then such fingerprint can be used to securely verify the user of interest. Thus, well-designed voice watermarking can help tandem systems in managing identity security in the voice inputs.

Considering honesty and inviolability as the first step approaching privacy preservation, voice signature is a worthy direction towards delivering voice integrity. However, current speech watermarking is designed for fixed-length offline audio files,~\eg, meeting recordings, and does not consider the impact of environmental conditions,~\eg, bitrate variability and background noise~\cite{zhang2021practical}. Such environmental factors make watermark embedding and retrieval very challenging. Further, they are not designed for real-time speaker recognition systems where input speech is unknown a priori and can be of variable length~\cite{zhang2021practical}. Therefore, speech watermarking must be extended to address the above challenges (\eg~how and which features to encode) for efficient practical applications.

\subsection{Trust vs Trustworthy}
Speech is a biometric characteristic of human beings, which can produce distinguishing and repeatable biometric features. Controversy has thus arisen over the risks of privacy and security around it.

\subsubsection{Is Conflict a Fundamental Principle?\\}

\emph{Privacy as Trust.} Should we suppress the speaker-related information including his/her identity for privacy preservation? Speaker-related information typically involves timbre, pitch, speaking rate, and speaking style. With the growth of advanced speech synthesis techniques, it is also easy to build speech synthesis systems (\ie~anonymization) from acquired data and then generate new speech samples which reflect the voice of a pseudo speaker. The genuine user can use the generated utterances for privacy protection against an automatic speaker verification (ASV) system. The hiding of speaker identity is also referred to as speaker anonymization or de-identification. These solutions propose to prevent access to the identity in order not to prevent improper use of it.

\emph{Security as Trustworthy.} Should we maintain the speaker-related information including his/her identity for security integrity? Voice-based authentication has been implemented in security-sensitive applications (\eg~smart home systems) to enable legitimate access. With the growth of advanced speech synthesis techniques, it is also easy to build speech synthesis systems (\ie~spoofing) from acquired data and then generate new speech samples which reflect the voice of a pseudo speaker. The adversary can use the generated utterances to attack an automatic speaker verification (ASV) system. The hiding of speaker identity is also referred to as speaker spoofing or presentation. These systems are used to gain illegitimate access with claimed identity to services protected by ASV.

We leave the question of deciding whether the trust-trustworthy conflict is fundamental (\ie~how to design the next generation of voice-based applications) as an open question for the research community.

\subsubsection{Beyond Voice Analytics.}
Our experiments so far focused on the speech processing domain. 
The development of synthesized techniques (\eg~generative models) are in every domain such as images, videos, etc., these tools have become a tough challenge. Recently, synthesized techniques are proposed in limiting the privacy risks by sharing synthetic data instead of real data in a manner that protects the privacy and preserves data utility~\cite{oprisanuevaluating, 9034117}. However, such techniques also can advance the development of deepfake techniques, depend on generating synthesized samples to attack the target systems. The problem might be expanding to become a broader question belong to~\textit{privacy and identity management}. Thus, there is an urgent need to develop countermeasures techniques against deepfake consequences. 

The need for trustworthy systems that offer end-to-end privacy guarantees is urgent~\cite{rogers2019privacy}. The importance of understanding and accommodating context (\ie~control over deployment/application) is a critical key behind designing privacy-preserving solutions to offer any degree of authenticity and linkability. To be considered privacy-enhancing, such a solution needs to allow the user to choose his required and acceptable degree of anonymization while maintaining the conventional capabilities for identification and authentication. 

%% file: related_work.tex
\section{Related Work}
\label{sec:related_work}

In this section, we overview the voice conversion technology in terms of its usage for privacy protection and security concerns against it.

\subsection{Voice Conversion}
Voice conversion is part of the general field of speech synthesis, where we convert text to speech or changes the properties of speech; for example, voice identity and emotion~\cite{9262021, aloufi2019emotionless}. VC tools modify speaker-dependent characteristics of the speech signal, such as spectral and prosodic aspects, while maintaining the speaker-independent information (\ie~linguistic). VC enables a wide range of applications including personalized speech synthesis~\cite{veaux2013towards, 9053104}, speaker de-identification~\cite{Hidebehind_2018, tomashenko2020introducing, 255304, 9247219}, and voice disguise~\cite{wang2020asvspoof}.

\textbf{Preserving Voice Privacy.}
VC mechanisms show their effectiveness in filtering out the speaker-related voice biometrics present in speech data without altering the linguistic content, thus preserving the usefulness of the shared data while protecting the users' privacy.  Most of the proposed works focus on protecting/anonymizing the speaker identity using these mechanisms~\cite{Hidebehind_2018, srivastava2019privacy, srivastava2020design}. For example, VoiceMask builds upon voice conversion to perturb the speech and then sends the sanitized speech audio to the voice input apps~\cite{Hidebehind_2018}. Similarly, Srivastava~\etal~in~\cite{srivastava2020design} propose an x-vector-based anonymization scheme to convert any input voice into a random pseudo-speaker based on the selected gender and region of x-vector space of the target pseudo-speaker. In~\cite{9247219} it is proposed an algorithm that produces anonymized speeches by adopting many-to-many voice conversion techniques based on variational autoencoders (VAEs) and modifying the speaker identity vectors of the VAE input to anonymize the speech data. Although these VC methods may provide some identity protection against less knowledgeable attackers (\ie~linkage attacks), they are unable to defend against an attacker that has extensive knowledge of the type of conversion and how it has been applied~\cite{9053868}.

Besides the speaker identity, various works have proposed to use VC-based mechanisms to protect a speaker's gender~\cite{jaiswal2019privacy} or emotions~\cite{emotionless_2019}. Champion~\etal~in~\cite{champion2021study} propose to alter other paralinguistic features (\ie~F0) and analyze the impact of this modification across gender. They found that the proposed F0 modification always improves pseudonymization, and both sources and target speaker genders affect the performance gain when modifying the F0. In~\cite{emotionless_2019}, an edge-based system is proposed to filter patterns from a user's voice before sharing it with cloud services for further analysis. Likewise, Vaidya~\etal~in~\cite{8844600} introduce an audio sanitizer, a software audio processor that filters and modifies the voice characteristics of the speaker from audio commands before they leave the client device by altering speech features (\ie~the short-term spectral features, spectro-temporal features, and high level features) in these commands.

\textbf{Voice Spoofing.}
VC poses a significant security threat wherever the voice is used as an authenticator~\cite{8844600}. VC has recently become one of the most easily accessible techniques to carry out spoofing attacks, presenting a threat to speaker verification systems (ASV). There are at least four major classes of spoofing attacks: impersonation, replay, speech synthesis, and voice conversion~\cite{marcel2019handbook}. The execution of speech synthesis and voice conversion attacks usually requires sophisticated speech technology. Speech synthesis systems can be used to generate entirely artificial speech signals, whereas voice conversion systems operate on natural speech~\cite{wu2014voice}. With sufficient training data, both speech synthesis and voice conversion technologies can produce high-quality speech signals that mimic the speech of a specific target speaker and are also highly effective in manipulating ASV systems. Such synthetic speech can be used to spoof the voice authentication systems and gain access to the user’s private resources (\eg~fraud attacks).

The awareness of this threat spawned research on anti-spoofing, including techniques to distinguish between bona fide and spoofed biometric data. Solutions are referred to as spoofing countermeasures or presentation attack detection systems~\cite{wang2020asvspoof}. For example, in~\cite{9198912}, an introduced approach to estimate the restoration function is proposed by minimizing a function of ASV scores to improve the defense against the automatic voice disguise (AVD) conducted by VC-based methods. Therefore, the improved conversion technologies also led to concerns about security and authentication. It is thus desirable to be able to prevent one's voice from being improperly used with such voice conversion technologies. 

\subsection{Privacy Exposure}
Deep learning has been a driving force in research and practice across speech application domains raising the need to study what causes privacy leaks and under which conditions a model is sensitive to different types of privacy-related attacks. In privacy-related attacks, the goal of an adversary is to gain knowledge that was not intended to be shared. Such knowledge can be about the training data or information about the model, or even extracting information about attributes of the data, such as unintentionally encoded information~\cite{8835245}.

\textbf{Membership Inference Attacks.}
In membership inference attacks (MIAs), the attacker aims to identify if a data record was used to train a machine learning model~\cite{8835245}. The attack is driven by the different behaviors of the target model when making predictions on samples within or out of its training set~\cite{chen2020gan, murakonda2020ml}. Song and Shmatikov~\cite{song2019auditing} discuss the application of user-level membership inference on text generative models, exploiting several top ranked outputs of the model. In the speech domain, Miao~\etal~in~\cite{miao2020audio} examine user-level membership inference (\ie~if this user has any data within target model’s training set) in the problem space of voice services, by designing an audio auditor to verify whether a specific user had unwillingly contributed audio used to train an automatic speech recognition (ASR) model under strict black-box access. Song~\etal~in~\cite{song2019privacy} combine the privacy and security domains by utilizing the success accuracy of membership inference attacks in reflecting the information leakage of training algorithms about individual members of the training set. 

\textbf{Reconstructing Attacks}
In a reconstruction attack, the attacker aims to infer attributes of the records in the training set~\cite{8835245, dwork2017exposed} by leveraging publicly accessible data that are not explicitly encoded as features or are not correlated to the learning task. `Overlearning' may cause revealing privacy- and bias-sensitive attributes that are not part of the target objective~\cite{song2020overlearning}. In the speech domain, it is possible to accurately infer a user's sensitive and private attributes (\eg~their emotion, sex, or health status) from deep acoustic models (\eg~DeepSpeech2). An attacker (\eg~a `curious' service provider) may use an acoustic model trained for speech recognition or speaker verification to learn further sensitive attributes from user input even if not present in its training data~\cite{aloufi2020privacypreserving}. Linkage attacks can be designed depending on the attackers’ knowledge about the anonymization scheme to infer the speaker’s identity~\cite{Srivastava_2019}. These types of attributes can lead to a secondary use that may include targeting content, or data brokers might profit from selling this information to other parties such as advertisers and insurance companies, or surveillance agencies may use these attributes to recognize users and track their activities and behaviors.

%% file: conclusion.tex
\section{Conclusion}
\label{sec:conclusion}

This work represents a step towards understanding the security risks of anonymization tools using a tandem evaluation framework. We show both empirically and analytically that (i) there exist intriguing effects between the two vector domains, (ii) an adversary can exploit these effects to optimize attacks with respect to multiple metrics, and (iii) it requires carefully accounting for such effects in designing effective countermeasures against the potential security and privacy attacks on VUI-based systems. We believe our findings shed light on the inherent vulnerabilities of VUIs deployed under realistic settings. This work also opens a few avenues for further investigation. Devising a unified evaluation framework accounting for both security and privacy may serve as a promising starting point for developing effective countermeasures. The detailed analysis in our paper highlights the importance of thinking about their combination.

%% file: CCS21.bbl

\begin{thebibliography}{92}


\ifx \showCODEN    \undefined \def \showCODEN     #1{\unskip}     \fi
\ifx \showDOI      \undefined \def \showDOI       #1{#1}\fi
\ifx \showISBNx    \undefined \def \showISBNx     #1{\unskip}     \fi
\ifx \showISBNxiii \undefined \def \showISBNxiii  #1{\unskip}     \fi
\ifx \showISSN     \undefined \def \showISSN      #1{\unskip}     \fi
\ifx \showLCCN     \undefined \def \showLCCN      #1{\unskip}     \fi
\ifx \shownote     \undefined \def \shownote      #1{#1}          \fi
\ifx \showarticletitle \undefined \def \showarticletitle #1{#1}   \fi
\ifx \showURL      \undefined \def \showURL       {\relax}        \fi
\providecommand\bibfield[2]{#2}
\providecommand\bibinfo[2]{#2}
\providecommand\natexlab[1]{#1}
\providecommand\showeprint[2][]{arXiv:#2}

\bibitem[\protect\citeauthoryear{Abdoli, Hafemann, Rony, Ayed, Cardinal, and
  Koerich}{Abdoli et~al\mbox{.}}{2019}]%
        {abdoli2019universal}
\bibfield{author}{\bibinfo{person}{Sajjad Abdoli}, \bibinfo{person}{Luiz~G
  Hafemann}, \bibinfo{person}{Jerome Rony}, \bibinfo{person}{Ismail~Ben Ayed},
  \bibinfo{person}{Patrick Cardinal}, {and} \bibinfo{person}{Alessandro~L
  Koerich}.} \bibinfo{year}{2019}\natexlab{}.
\newblock \showarticletitle{Universal adversarial audio perturbations}.
\newblock \bibinfo{journal}{\emph{arXiv preprint arXiv:1908.03173}}
  (\bibinfo{year}{2019}).
\newblock


\bibitem[\protect\citeauthoryear{Abdullah, Garcia, Peeters, Traynor, Butler,
  and Wilson}{Abdullah et~al\mbox{.}}{2019a}]%
        {abdullah2019practical}
\bibfield{author}{\bibinfo{person}{Hadi Abdullah}, \bibinfo{person}{Washington
  Garcia}, \bibinfo{person}{Christian Peeters}, \bibinfo{person}{Patrick
  Traynor}, \bibinfo{person}{Kevin~RB Butler}, {and} \bibinfo{person}{Joseph
  Wilson}.} \bibinfo{year}{2019}\natexlab{a}.
\newblock \showarticletitle{Practical hidden voice attacks against speech and
  speaker recognition systems}.
\newblock \bibinfo{journal}{\emph{arXiv preprint arXiv:1904.05734}}
  (\bibinfo{year}{2019}).
\newblock


\bibitem[\protect\citeauthoryear{Abdullah, Rahman, Garcia, Blue, Warren, Yadav,
  Shrimpton, and Traynor}{Abdullah et~al\mbox{.}}{2019b}]%
        {abdullah2019hear}
\bibfield{author}{\bibinfo{person}{Hadi Abdullah},
  \bibinfo{person}{Muhammad~Sajidur Rahman}, \bibinfo{person}{Washington
  Garcia}, \bibinfo{person}{Logan Blue}, \bibinfo{person}{Kevin Warren},
  \bibinfo{person}{Anurag~Swarnim Yadav}, \bibinfo{person}{Tom Shrimpton},
  {and} \bibinfo{person}{Patrick Traynor}.} \bibinfo{year}{2019}\natexlab{b}.
\newblock \showarticletitle{Hear" no evil", see" kenansville": Efficient and
  transferable black-box attacks on speech recognition and voice identification
  systems}.
\newblock \bibinfo{journal}{\emph{arXiv preprint arXiv:1910.05262}}
  (\bibinfo{year}{2019}).
\newblock


\bibitem[\protect\citeauthoryear{Ahmed, Chowdhury, Fawaz, and Ramanathan}{Ahmed
  et~al\mbox{.}}{2020}]%
        {255304}
\bibfield{author}{\bibinfo{person}{Shimaa Ahmed}, \bibinfo{person}{Amrita~Roy
  Chowdhury}, \bibinfo{person}{Kassem Fawaz}, {and} \bibinfo{person}{Parmesh
  Ramanathan}.} \bibinfo{year}{2020}\natexlab{}.
\newblock \showarticletitle{Preech: A System for Privacy-Preserving Speech
  Transcription}. In \bibinfo{booktitle}{\emph{29th {USENIX} Security Symposium
  ({USENIX} Security 20)}}. \bibinfo{publisher}{{USENIX} Association},
  \bibinfo{pages}{2703--2720}.
\newblock
\showISBNx{978-1-939133-17-5}
\urldef\tempurl%
\url{https://www.usenix.org/conference/usenixsecurity20/presentation/ahmed-shimaa}
\showURL{%
\tempurl}


\bibitem[\protect\citeauthoryear{Aloufi, Haddadi, and Boyle}{Aloufi
  et~al\mbox{.}}{2019a}]%
        {emotionless_2019}
\bibfield{author}{\bibinfo{person}{Ranya Aloufi}, \bibinfo{person}{Hamed
  Haddadi}, {and} \bibinfo{person}{David Boyle}.}
  \bibinfo{year}{2019}\natexlab{a}.
\newblock \showarticletitle{Emotion Filtering at the Edge}. In
  \bibinfo{booktitle}{\emph{Proceedings of the 1st Workshop on Machine Learning
  on Edge in Sensor Systems}} (New York, NY, USA).
  \bibinfo{publisher}{Association for Computing Machinery}.
\newblock
\urldef\tempurl%
\url{https://doi.org/10.1145/3362743.3362960}
\showDOI{\tempurl}


\bibitem[\protect\citeauthoryear{Aloufi, Haddadi, and Boyle}{Aloufi
  et~al\mbox{.}}{2019b}]%
        {aloufi2019emotionless}
\bibfield{author}{\bibinfo{person}{Ranya Aloufi}, \bibinfo{person}{Hamed
  Haddadi}, {and} \bibinfo{person}{David Boyle}.}
  \bibinfo{year}{2019}\natexlab{b}.
\newblock \bibinfo{title}{Emotionless: Privacy-Preserving Speech Analysis for
  Voice Assistants}.
\newblock
\newblock
\showeprint[arxiv]{1908.03632}~[cs.CR]


\bibitem[\protect\citeauthoryear{Aloufi, Haddadi, and Boyle}{Aloufi
  et~al\mbox{.}}{2020}]%
        {aloufi2020privacypreserving}
\bibfield{author}{\bibinfo{person}{Ranya Aloufi}, \bibinfo{person}{Hamed
  Haddadi}, {and} \bibinfo{person}{David Boyle}.}
  \bibinfo{year}{2020}\natexlab{}.
\newblock \bibinfo{booktitle}{\emph{Privacy-Preserving Voice Analysis via
  Disentangled Representations}}.
\newblock \bibinfo{publisher}{Association for Computing Machinery},
  \bibinfo{address}{New York, NY, USA}, \bibinfo{pages}{1–14}.
\newblock
\showISBNx{9781450380843}
\urldef\tempurl%
\url{https://doi.org/10.1145/3411495.3421355}
\showURL{%
\tempurl}


\bibitem[\protect\citeauthoryear{Aloufi, Haddadi, and Boyle}{Aloufi
  et~al\mbox{.}}{2021}]%
        {aloufi2021configurable}
\bibfield{author}{\bibinfo{person}{Ranya Aloufi}, \bibinfo{person}{Hamed
  Haddadi}, {and} \bibinfo{person}{David Boyle}.}
  \bibinfo{year}{2021}\natexlab{}.
\newblock \showarticletitle{Configurable Privacy-Preserving Automatic Speech
  Recognition}.
\newblock \bibinfo{journal}{\emph{arXiv preprint arXiv:2104.00766}}
  (\bibinfo{year}{2021}).
\newblock


\bibitem[\protect\citeauthoryear{Alzantot, Balaji, and Srivastava}{Alzantot
  et~al\mbox{.}}{2018}]%
        {alzantot2018did}
\bibfield{author}{\bibinfo{person}{Moustafa Alzantot},
  \bibinfo{person}{Bharathan Balaji}, {and} \bibinfo{person}{Mani Srivastava}.}
  \bibinfo{year}{2018}\natexlab{}.
\newblock \showarticletitle{Did you hear that? adversarial examples against
  automatic speech recognition}.
\newblock \bibinfo{journal}{\emph{arXiv preprint arXiv:1801.00554}}
  (\bibinfo{year}{2018}).
\newblock


\bibitem[\protect\citeauthoryear{Bhattacharya, Alam, Stafylakis, and
  Kenny}{Bhattacharya et~al\mbox{.}}{2016}]%
        {bhattacharya2016deep}
\bibfield{author}{\bibinfo{person}{Gautam Bhattacharya},
  \bibinfo{person}{Jahangir Alam}, \bibinfo{person}{Themos Stafylakis}, {and}
  \bibinfo{person}{Patrick Kenny}.} \bibinfo{year}{2016}\natexlab{}.
\newblock \showarticletitle{Deep neural network based text-dependent speaker
  recognition: Preliminary results}. In \bibinfo{booktitle}{\emph{Proc.
  Odyssey}}. \bibinfo{pages}{2--15}.
\newblock


\bibitem[\protect\citeauthoryear{Carlini and Wagner}{Carlini and
  Wagner}{2018}]%
        {carlini2018audio}
\bibfield{author}{\bibinfo{person}{Nicholas Carlini} {and}
  \bibinfo{person}{David Wagner}.} \bibinfo{year}{2018}\natexlab{}.
\newblock \showarticletitle{Audio adversarial examples: Targeted attacks on
  speech-to-text}. In \bibinfo{booktitle}{\emph{2018 IEEE Security and Privacy
  Workshops (SPW)}}. IEEE, \bibinfo{pages}{1--7}.
\newblock


\bibitem[\protect\citeauthoryear{Champion, Jouvet, and Larcher}{Champion
  et~al\mbox{.}}{2021}]%
        {champion2021study}
\bibfield{author}{\bibinfo{person}{Pierre Champion}, \bibinfo{person}{Denis
  Jouvet}, {and} \bibinfo{person}{Anthony Larcher}.}
  \bibinfo{year}{2021}\natexlab{}.
\newblock \bibinfo{title}{A Study of F0 Modification for X-Vector Based Speech
  Pseudonymization Across Gender}.
\newblock
\newblock
\showeprint[arxiv]{2101.08478}~[eess.AS]


\bibitem[\protect\citeauthoryear{Chen, Yu, Zhang, and Fritz}{Chen
  et~al\mbox{.}}{2020}]%
        {chen2020gan}
\bibfield{author}{\bibinfo{person}{Dingfan Chen}, \bibinfo{person}{Ning Yu},
  \bibinfo{person}{Yang Zhang}, {and} \bibinfo{person}{Mario Fritz}.}
  \bibinfo{year}{2020}\natexlab{}.
\newblock \showarticletitle{Gan-leaks: A taxonomy of membership inference
  attacks against generative models}. In \bibinfo{booktitle}{\emph{Proceedings
  of the 2020 ACM SIGSAC Conference on Computer and Communications Security}}.
  \bibinfo{pages}{343--362}.
\newblock


\bibitem[\protect\citeauthoryear{Chen, Qian, and Yu}{Chen
  et~al\mbox{.}}{2015}]%
        {chen2015multi}
\bibfield{author}{\bibinfo{person}{Nanxin Chen}, \bibinfo{person}{Yanmin Qian},
  {and} \bibinfo{person}{Kai Yu}.} \bibinfo{year}{2015}\natexlab{}.
\newblock \showarticletitle{Multi-task learning for text-dependent speaker
  verification}. In \bibinfo{booktitle}{\emph{Sixteenth annual conference of
  the international speech communication association}}.
\newblock


\bibitem[\protect\citeauthoryear{Chettri, Kinnunen, and Benetos}{Chettri
  et~al\mbox{.}}{2020}]%
        {chettri2020subband}
\bibfield{author}{\bibinfo{person}{Bhusan Chettri}, \bibinfo{person}{Tomi
  Kinnunen}, {and} \bibinfo{person}{Emmanouil Benetos}.}
  \bibinfo{year}{2020}\natexlab{}.
\newblock \showarticletitle{Subband modeling for spoofing detection in
  automatic speaker verification}.
\newblock \bibinfo{journal}{\emph{arXiv preprint arXiv:2004.01922}}
  (\bibinfo{year}{2020}).
\newblock


\bibitem[\protect\citeauthoryear{Cisse, Adi, Neverova, and Keshet}{Cisse
  et~al\mbox{.}}{2017}]%
        {cisse2017houdini}
\bibfield{author}{\bibinfo{person}{Moustapha~M Cisse}, \bibinfo{person}{Yossi
  Adi}, \bibinfo{person}{Natalia Neverova}, {and} \bibinfo{person}{Joseph
  Keshet}.} \bibinfo{year}{2017}\natexlab{}.
\newblock \showarticletitle{Houdini: Fooling deep structured visual and speech
  recognition models with adversarial examples}.
\newblock \bibinfo{journal}{\emph{Advances in neural information processing
  systems}}  \bibinfo{volume}{30} (\bibinfo{year}{2017}).
\newblock


\bibitem[\protect\citeauthoryear{Das, Tian, Kinnunen, and Li}{Das
  et~al\mbox{.}}{2020}]%
        {das2020attacker}
\bibfield{author}{\bibinfo{person}{Rohan~Kumar Das}, \bibinfo{person}{Xiaohai
  Tian}, \bibinfo{person}{Tomi Kinnunen}, {and} \bibinfo{person}{Haizhou Li}.}
  \bibinfo{year}{2020}\natexlab{}.
\newblock \showarticletitle{The Attacker's Perspective on Automatic Speaker
  Verification: An Overview}.
\newblock \bibinfo{journal}{\emph{arXiv preprint arXiv:2004.08849}}
  (\bibinfo{year}{2020}).
\newblock


\bibitem[\protect\citeauthoryear{{Das}, {Yang}, and {Li}}{{Das}
  et~al\mbox{.}}{2020}]%
        {9053086}
\bibfield{author}{\bibinfo{person}{R.~K. {Das}}, \bibinfo{person}{J. {Yang}},
  {and} \bibinfo{person}{H. {Li}}.} \bibinfo{year}{2020}\natexlab{}.
\newblock \showarticletitle{Assessing the Scope of Generalized Countermeasures
  for Anti-Spoofing}. In \bibinfo{booktitle}{\emph{ICASSP 2020 - 2020 IEEE
  International Conference on Acoustics, Speech and Signal Processing
  (ICASSP)}}. \bibinfo{pages}{6589--6593}.
\newblock
\urldef\tempurl%
\url{https://doi.org/10.1109/ICASSP40776.2020.9053086}
\showDOI{\tempurl}


\bibitem[\protect\citeauthoryear{Dehak, Kenny, Dehak, Dumouchel, and
  Ouellet}{Dehak et~al\mbox{.}}{2010}]%
        {dehak2010front}
\bibfield{author}{\bibinfo{person}{Najim Dehak}, \bibinfo{person}{Patrick~J
  Kenny}, \bibinfo{person}{R{\'e}da Dehak}, \bibinfo{person}{Pierre Dumouchel},
  {and} \bibinfo{person}{Pierre Ouellet}.} \bibinfo{year}{2010}\natexlab{}.
\newblock \showarticletitle{Front-end factor analysis for speaker
  verification}.
\newblock \bibinfo{journal}{\emph{IEEE Transactions on Audio, Speech, and
  Language Processing}} \bibinfo{volume}{19}, \bibinfo{number}{4}
  (\bibinfo{year}{2010}), \bibinfo{pages}{788--798}.
\newblock


\bibitem[\protect\citeauthoryear{Dubois, Kolcun, Mandalari, Paracha, Choffnes,
  and Haddadi}{Dubois et~al\mbox{.}}{2020}]%
        {PETS_2020smarthome}
\bibfield{author}{\bibinfo{person}{Daniel~J. Dubois}, \bibinfo{person}{Roman
  Kolcun}, \bibinfo{person}{Anna~Maria Mandalari},
  \bibinfo{person}{Muhammad~Talha Paracha}, \bibinfo{person}{David Choffnes},
  {and} \bibinfo{person}{Hamed Haddadi}.} \bibinfo{year}{2020}\natexlab{}.
\newblock \showarticletitle{When Speakers Are All Ears: Characterizing
  Misactivations of IoT Smart Speakers}. In
  \bibinfo{booktitle}{\emph{Proceedings of the 20th Privacy Enhancing
  Technologies Symposium (PETS 2020)}} (Montreal, Canada).
\newblock


\bibitem[\protect\citeauthoryear{Dwork, Smith, Steinke, and Ullman}{Dwork
  et~al\mbox{.}}{2017}]%
        {dwork2017exposed}
\bibfield{author}{\bibinfo{person}{Cynthia Dwork}, \bibinfo{person}{Adam
  Smith}, \bibinfo{person}{Thomas Steinke}, {and} \bibinfo{person}{Jonathan
  Ullman}.} \bibinfo{year}{2017}\natexlab{}.
\newblock \showarticletitle{Exposed! a survey of attacks on private
  data.(2017)}.
\newblock  (\bibinfo{year}{2017}).
\newblock


\bibitem[\protect\citeauthoryear{Fu, Qian, Liu, and Yu}{Fu
  et~al\mbox{.}}{2014}]%
        {fu2014tandem}
\bibfield{author}{\bibinfo{person}{Tianfan Fu}, \bibinfo{person}{Yanmin Qian},
  \bibinfo{person}{Yuan Liu}, {and} \bibinfo{person}{Kai Yu}.}
  \bibinfo{year}{2014}\natexlab{}.
\newblock \showarticletitle{Tandem deep features for text-dependent speaker
  verification}. In \bibinfo{booktitle}{\emph{Fifteenth Annual Conference of
  the International Speech Communication Association}}.
\newblock


\bibitem[\protect\citeauthoryear{Goodfellow, Shlens, and Szegedy}{Goodfellow
  et~al\mbox{.}}{2014}]%
        {goodfellow2014explaining}
\bibfield{author}{\bibinfo{person}{Ian~J Goodfellow}, \bibinfo{person}{Jonathon
  Shlens}, {and} \bibinfo{person}{Christian Szegedy}.}
  \bibinfo{year}{2014}\natexlab{}.
\newblock \showarticletitle{Explaining and harnessing adversarial examples}.
\newblock \bibinfo{journal}{\emph{arXiv preprint arXiv:1412.6572}}
  (\bibinfo{year}{2014}).
\newblock


\bibitem[\protect\citeauthoryear{Griffin and Lim}{Griffin and Lim}{1984}]%
        {griffin1984signal}
\bibfield{author}{\bibinfo{person}{Daniel Griffin} {and} \bibinfo{person}{Jae
  Lim}.} \bibinfo{year}{1984}\natexlab{}.
\newblock \showarticletitle{Signal estimation from modified short-time Fourier
  transform}.
\newblock \bibinfo{journal}{\emph{IEEE Transactions on Acoustics, Speech, and
  Signal Processing}} \bibinfo{volume}{32}, \bibinfo{number}{2}
  (\bibinfo{year}{1984}), \bibinfo{pages}{236--243}.
\newblock


\bibitem[\protect\citeauthoryear{{Han}, {Li}, {Cao}, {Ma}, and
  {Yoshikawa}}{{Han} et~al\mbox{.}}{2020}]%
        {9102875}
\bibfield{author}{\bibinfo{person}{Y. {Han}}, \bibinfo{person}{S. {Li}},
  \bibinfo{person}{Y. {Cao}}, \bibinfo{person}{Q. {Ma}}, {and}
  \bibinfo{person}{M. {Yoshikawa}}.} \bibinfo{year}{2020}\natexlab{}.
\newblock \showarticletitle{Voice-Indistinguishability: Protecting Voiceprint
  In Privacy-Preserving Speech Data Release}. In \bibinfo{booktitle}{\emph{2020
  IEEE International Conference on Multimedia and Expo (ICME)}}.
  \bibinfo{pages}{1--6}.
\newblock
\urldef\tempurl%
\url{https://doi.org/10.1109/ICME46284.2020.9102875}
\showDOI{\tempurl}


\bibitem[\protect\citeauthoryear{Heigold, Moreno, Bengio, and Shazeer}{Heigold
  et~al\mbox{.}}{2016}]%
        {heigold2016end}
\bibfield{author}{\bibinfo{person}{Georg Heigold}, \bibinfo{person}{Ignacio
  Moreno}, \bibinfo{person}{Samy Bengio}, {and} \bibinfo{person}{Noam
  Shazeer}.} \bibinfo{year}{2016}\natexlab{}.
\newblock \showarticletitle{End-to-end text-dependent speaker verification}. In
  \bibinfo{booktitle}{\emph{2016 IEEE International Conference on Acoustics,
  Speech and Signal Processing (ICASSP)}}. IEEE, \bibinfo{pages}{5115--5119}.
\newblock


\bibitem[\protect\citeauthoryear{Hemavathi and Kumaraswamy}{Hemavathi and
  Kumaraswamy}{2021}]%
        {hemavathi2021voice}
\bibfield{author}{\bibinfo{person}{R Hemavathi} {and} \bibinfo{person}{R
  Kumaraswamy}.} \bibinfo{year}{2021}\natexlab{}.
\newblock \showarticletitle{Voice conversion spoofing detection by exploring
  artifacts estimates}.
\newblock \bibinfo{journal}{\emph{Multimedia Tools and Applications}}
  (\bibinfo{year}{2021}), \bibinfo{pages}{1--20}.
\newblock


\bibitem[\protect\citeauthoryear{Ho and Akagi}{Ho and Akagi}{2021}]%
        {9367139}
\bibfield{author}{\bibinfo{person}{Tuan~Vu Ho} {and} \bibinfo{person}{Masato
  Akagi}.} \bibinfo{year}{2021}\natexlab{}.
\newblock \showarticletitle{Cross-Lingual Voice Conversion With Controllable
  Speaker Individuality Using Variational Autoencoder and Star Generative
  Adversarial Network}.
\newblock \bibinfo{journal}{\emph{IEEE Access}}  \bibinfo{volume}{9}
  (\bibinfo{year}{2021}), \bibinfo{pages}{47503--47515}.
\newblock
\urldef\tempurl%
\url{https://doi.org/10.1109/ACCESS.2021.3063519}
\showDOI{\tempurl}


\bibitem[\protect\citeauthoryear{Hu, Chou, and Lee}{Hu et~al\mbox{.}}{2021}]%
        {9316175}
\bibfield{author}{\bibinfo{person}{Hwai-Tsu Hu}, \bibinfo{person}{Hsien-Hsin
  Chou}, {and} \bibinfo{person}{Tung-Tsun Lee}.}
  \bibinfo{year}{2021}\natexlab{}.
\newblock \showarticletitle{Robust Blind Speech Watermarking via FFT-Based
  Perceptual Vector Norm Modulation With Frame Self-Synchronization}.
\newblock \bibinfo{journal}{\emph{IEEE Access}}  \bibinfo{volume}{9}
  (\bibinfo{year}{2021}), \bibinfo{pages}{9916--9925}.
\newblock
\urldef\tempurl%
\url{https://doi.org/10.1109/ACCESS.2021.3049525}
\showDOI{\tempurl}


\bibitem[\protect\citeauthoryear{{Huang}, {He}, {Wei}, {Gale}, {Li}, and
  {Gong}}{{Huang} et~al\mbox{.}}{2020}]%
        {9053104}
\bibfield{author}{\bibinfo{person}{Y. {Huang}}, \bibinfo{person}{L. {He}},
  \bibinfo{person}{W. {Wei}}, \bibinfo{person}{W. {Gale}}, \bibinfo{person}{J.
  {Li}}, {and} \bibinfo{person}{Y. {Gong}}.} \bibinfo{year}{2020}\natexlab{}.
\newblock \showarticletitle{Using Personalized Speech Synthesis and Neural
  Language Generator for Rapid Speaker Adaptation}. In
  \bibinfo{booktitle}{\emph{ICASSP 2020 - 2020 IEEE International Conference on
  Acoustics, Speech and Signal Processing (ICASSP)}}.
  \bibinfo{pages}{7399--7403}.
\newblock
\urldef\tempurl%
\url{https://doi.org/10.1109/ICASSP40776.2020.9053104}
\showDOI{\tempurl}


\bibitem[\protect\citeauthoryear{Jaiswal and Provost}{Jaiswal and
  Provost}{2019}]%
        {jaiswal2019privacy}
\bibfield{author}{\bibinfo{person}{Mimansa Jaiswal} {and}
  \bibinfo{person}{Emily~Mower Provost}.} \bibinfo{year}{2019}\natexlab{}.
\newblock \bibinfo{title}{Privacy Enhanced Multimodal Neural Representations
  for Emotion Recognition}.
\newblock
\newblock
\showeprint[arxiv]{1910.13212}~[cs.LG]


\bibitem[\protect\citeauthoryear{Kalchbrenner, Elsen, Simonyan, Noury,
  Casagrande, Lockhart, Stimberg, van~den Oord, Dieleman, and
  Kavukcuoglu}{Kalchbrenner et~al\mbox{.}}{2018}]%
        {kalchbrenner2018efficient}
\bibfield{author}{\bibinfo{person}{Nal Kalchbrenner}, \bibinfo{person}{Erich
  Elsen}, \bibinfo{person}{Karen Simonyan}, \bibinfo{person}{Seb Noury},
  \bibinfo{person}{Norman Casagrande}, \bibinfo{person}{Edward Lockhart},
  \bibinfo{person}{Florian Stimberg}, \bibinfo{person}{Aaron van~den Oord},
  \bibinfo{person}{Sander Dieleman}, {and} \bibinfo{person}{Koray
  Kavukcuoglu}.} \bibinfo{year}{2018}\natexlab{}.
\newblock \showarticletitle{Efficient Neural Audio Synthesis}. In
  \bibinfo{booktitle}{\emph{Proceedings of the 35th International Conference on
  Machine Learning}} \emph{(\bibinfo{series}{Proceedings of Machine Learning
  Research}, Vol.~\bibinfo{volume}{80})},
  \bibfield{editor}{\bibinfo{person}{Jennifer Dy} {and}
  \bibinfo{person}{Andreas Krause}} (Eds.). \bibinfo{publisher}{PMLR},
  \bibinfo{pages}{2410--2419}.
\newblock


\bibitem[\protect\citeauthoryear{Kawahara}{Kawahara}{2006}]%
        {kawahara2006straight}
\bibfield{author}{\bibinfo{person}{Hideki Kawahara}.}
  \bibinfo{year}{2006}\natexlab{}.
\newblock \showarticletitle{STRAIGHT, exploitation of the other aspect of
  VOCODER: Perceptually isomorphic decomposition of speech sounds}.
\newblock \bibinfo{journal}{\emph{Acoustical science and technology}}
  \bibinfo{volume}{27}, \bibinfo{number}{6} (\bibinfo{year}{2006}),
  \bibinfo{pages}{349--353}.
\newblock


\bibitem[\protect\citeauthoryear{Kinnunen, Delgado, Evans, Lee, Vestman,
  Nautsch, Todisco, Wang, Sahidullah, Yamagishi, et~al\mbox{.}}{Kinnunen
  et~al\mbox{.}}{2020}]%
        {kinnunen2020tandem}
\bibfield{author}{\bibinfo{person}{Tomi Kinnunen}, \bibinfo{person}{H{\'e}ctor
  Delgado}, \bibinfo{person}{Nicholas Evans}, \bibinfo{person}{Kong~Aik Lee},
  \bibinfo{person}{Ville Vestman}, \bibinfo{person}{Andreas Nautsch},
  \bibinfo{person}{Massimiliano Todisco}, \bibinfo{person}{Xin Wang},
  \bibinfo{person}{Md Sahidullah}, \bibinfo{person}{Junichi Yamagishi},
  {et~al\mbox{.}}} \bibinfo{year}{2020}\natexlab{}.
\newblock \showarticletitle{Tandem assessment of spoofing countermeasures and
  automatic speaker verification: Fundamentals}.
\newblock \bibinfo{journal}{\emph{IEEE/ACM Transactions on Audio, Speech, and
  Language Processing}}  \bibinfo{volume}{28} (\bibinfo{year}{2020}),
  \bibinfo{pages}{2195--2210}.
\newblock


\bibitem[\protect\citeauthoryear{Kobayashi, Huang, Wu, Tobing, Hayashi, and
  Toda}{Kobayashi et~al\mbox{.}}{2021}]%
        {kobayashi2021crank}
\bibfield{author}{\bibinfo{person}{Kazuhiro Kobayashi},
  \bibinfo{person}{Wen-Chin Huang}, \bibinfo{person}{Yi-Chiao Wu},
  \bibinfo{person}{Patrick~Lumban Tobing}, \bibinfo{person}{Tomoki Hayashi},
  {and} \bibinfo{person}{Tomoki Toda}.} \bibinfo{year}{2021}\natexlab{}.
\newblock \bibinfo{title}{crank: An Open-Source Software for Nonparallel Voice
  Conversion Based on Vector-Quantized Variational Autoencoder}.
\newblock
\newblock
\showeprint[arxiv]{2103.02858}~[eess.AS]


\bibitem[\protect\citeauthoryear{{Lal Srivastava}, {Vauquier}, {Sahidullah},
  {Bellet}, {Tommasi}, and {Vincent}}{{Lal Srivastava} et~al\mbox{.}}{2020}]%
        {9053868}
\bibfield{author}{\bibinfo{person}{B.~M. {Lal Srivastava}}, \bibinfo{person}{N.
  {Vauquier}}, \bibinfo{person}{M. {Sahidullah}}, \bibinfo{person}{A.
  {Bellet}}, \bibinfo{person}{M. {Tommasi}}, {and} \bibinfo{person}{E.
  {Vincent}}.} \bibinfo{year}{2020}\natexlab{}.
\newblock \showarticletitle{Evaluating Voice Conversion-Based Privacy
  Protection against Informed Attackers}. In \bibinfo{booktitle}{\emph{ICASSP
  2020 - 2020 IEEE International Conference on Acoustics, Speech and Signal
  Processing (ICASSP)}}. \bibinfo{pages}{2802--2806}.
\newblock
\urldef\tempurl%
\url{https://doi.org/10.1109/ICASSP40776.2020.9053868}
\showDOI{\tempurl}


\bibitem[\protect\citeauthoryear{Leeuwen and Br\"{u}mmer}{Leeuwen and
  Br\"{u}mmer}{2007}]%
        {cll}
\bibfield{author}{\bibinfo{person}{David~A. Leeuwen} {and}
  \bibinfo{person}{Niko Br\"{u}mmer}.} \bibinfo{year}{2007}\natexlab{}.
\newblock \bibinfo{booktitle}{\emph{An Introduction to Application-Independent
  Evaluation of Speaker Recognition Systems}}.
\newblock \bibinfo{publisher}{Springer-Verlag}, \bibinfo{address}{Berlin,
  Heidelberg}, \bibinfo{pages}{330–353}.
\newblock
\showISBNx{9783540741862}
\urldef\tempurl%
\url{https://doi.org/10.1007/978-3-540-74200-5_19}
\showURL{%
\tempurl}


\bibitem[\protect\citeauthoryear{Liu, Wu, Lee, and Meng}{Liu
  et~al\mbox{.}}{2019}]%
        {liu2019adversarial}
\bibfield{author}{\bibinfo{person}{Songxiang Liu}, \bibinfo{person}{Haibin Wu},
  \bibinfo{person}{Hung-yi Lee}, {and} \bibinfo{person}{Helen Meng}.}
  \bibinfo{year}{2019}\natexlab{}.
\newblock \showarticletitle{Adversarial attacks on spoofing countermeasures of
  automatic speaker verification}. In \bibinfo{booktitle}{\emph{2019 IEEE
  Automatic Speech Recognition and Understanding Workshop (ASRU)}}. IEEE,
  \bibinfo{pages}{312--319}.
\newblock


\bibitem[\protect\citeauthoryear{Liu, Qian, Chen, Fu, Zhang, and Yu}{Liu
  et~al\mbox{.}}{2015}]%
        {liu2015deep}
\bibfield{author}{\bibinfo{person}{Yuan Liu}, \bibinfo{person}{Yanmin Qian},
  \bibinfo{person}{Nanxin Chen}, \bibinfo{person}{Tianfan Fu},
  \bibinfo{person}{Ya Zhang}, {and} \bibinfo{person}{Kai Yu}.}
  \bibinfo{year}{2015}\natexlab{}.
\newblock \showarticletitle{Deep feature for text-dependent speaker
  verification}.
\newblock \bibinfo{journal}{\emph{Speech Communication}}  \bibinfo{volume}{73}
  (\bibinfo{year}{2015}), \bibinfo{pages}{1--13}.
\newblock


\bibitem[\protect\citeauthoryear{Lorenzo-Trueba, Drugman, Latorre, Merritt,
  Putrycz, Barra-Chicote, Moinet, and Aggarwal}{Lorenzo-Trueba
  et~al\mbox{.}}{2019}]%
        {lorenzo2018towards}
\bibfield{author}{\bibinfo{person}{Jaime Lorenzo-Trueba},
  \bibinfo{person}{Thomas Drugman}, \bibinfo{person}{Javier Latorre},
  \bibinfo{person}{Thomas Merritt}, \bibinfo{person}{Bartosz Putrycz},
  \bibinfo{person}{Roberto Barra-Chicote}, \bibinfo{person}{Alexis Moinet},
  {and} \bibinfo{person}{Vatsal Aggarwal}.} \bibinfo{year}{2019}\natexlab{}.
\newblock \showarticletitle{Towards Achieving Robust Universal Neural
  Vocoding}. \bibinfo{pages}{181--185}.
\newblock
\urldef\tempurl%
\url{https://doi.org/10.21437/Interspeech.2019-1424}
\showDOI{\tempurl}


\bibitem[\protect\citeauthoryear{Marcel, Nixon, Fierrez, and Evans}{Marcel
  et~al\mbox{.}}{2019}]%
        {marcel2019handbook}
\bibfield{author}{\bibinfo{person}{S{\'e}bastien Marcel},
  \bibinfo{person}{Mark~S Nixon}, \bibinfo{person}{Julian Fierrez}, {and}
  \bibinfo{person}{Nicholas Evans}.} \bibinfo{year}{2019}\natexlab{}.
\newblock \bibinfo{booktitle}{\emph{Handbook of biometric anti-spoofing:
  Presentation attack detection}}.
\newblock \bibinfo{publisher}{Springer}.
\newblock


\bibitem[\protect\citeauthoryear{Miao, Xue, Chen, Pan, Zhang, Zhao, Kaafar, and
  Xiang}{Miao et~al\mbox{.}}{2020}]%
        {miao2020audio}
\bibfield{author}{\bibinfo{person}{Yuantian Miao}, \bibinfo{person}{Minhui
  Xue}, \bibinfo{person}{Chao Chen}, \bibinfo{person}{Lei Pan},
  \bibinfo{person}{Jun Zhang}, \bibinfo{person}{Benjamin Zi~Hao Zhao},
  \bibinfo{person}{Dali Kaafar}, {and} \bibinfo{person}{Yang Xiang}.}
  \bibinfo{year}{2020}\natexlab{}.
\newblock \bibinfo{title}{The Audio Auditor: User-Level Membership Inference in
  Internet of Things Voice Services}.
\newblock
\newblock
\showeprint[arxiv]{1905.07082}~[cs.CR]


\bibitem[\protect\citeauthoryear{Morise, Yokomori, and Ozawa}{Morise
  et~al\mbox{.}}{2016}]%
        {morise2016world}
\bibfield{author}{\bibinfo{person}{Masanori Morise}, \bibinfo{person}{Fumiya
  Yokomori}, {and} \bibinfo{person}{Kenji Ozawa}.}
  \bibinfo{year}{2016}\natexlab{}.
\newblock \showarticletitle{WORLD: a vocoder-based high-quality speech
  synthesis system for real-time applications}.
\newblock \bibinfo{journal}{\emph{IEICE TRANSACTIONS on Information and
  Systems}} (\bibinfo{year}{2016}).
\newblock


\bibitem[\protect\citeauthoryear{Murakonda and Shokri}{Murakonda and
  Shokri}{2020}]%
        {murakonda2020ml}
\bibfield{author}{\bibinfo{person}{Sasi~Kumar Murakonda} {and}
  \bibinfo{person}{Reza Shokri}.} \bibinfo{year}{2020}\natexlab{}.
\newblock \showarticletitle{Ml privacy meter: Aiding regulatory compliance by
  quantifying the privacy risks of machine learning}.
\newblock \bibinfo{journal}{\emph{arXiv preprint arXiv:2007.09339}}
  (\bibinfo{year}{2020}).
\newblock


\bibitem[\protect\citeauthoryear{Nagrani, Chung, and Zisserman}{Nagrani
  et~al\mbox{.}}{2017}]%
        {nagrani2017voxceleb}
\bibfield{author}{\bibinfo{person}{Arsha Nagrani}, \bibinfo{person}{Joon~Son
  Chung}, {and} \bibinfo{person}{Andrew Zisserman}.}
  \bibinfo{year}{2017}\natexlab{}.
\newblock \showarticletitle{VoxCeleb: {A} Large-Scale Speaker Identification
  Dataset}. In \bibinfo{booktitle}{\emph{Interspeech 2017, 18th Annual
  Conference of the International Speech Communication Association, Stockholm,
  Sweden, August 20-24, 2017}}, \bibfield{editor}{\bibinfo{person}{Francisco
  Lacerda}} (Ed.). \bibinfo{publisher}{{ISCA}}, \bibinfo{pages}{2616--2620}.
\newblock


\bibitem[\protect\citeauthoryear{Nakamura, Saito, Takamichi, Ijima, and
  Saruwatari}{Nakamura et~al\mbox{.}}{2019}]%
        {nakamura2019v2s}
\bibfield{author}{\bibinfo{person}{Taiki Nakamura}, \bibinfo{person}{Yuki
  Saito}, \bibinfo{person}{Shinnosuke Takamichi}, \bibinfo{person}{Yusuke
  Ijima}, {and} \bibinfo{person}{Hiroshi Saruwatari}.}
  \bibinfo{year}{2019}\natexlab{}.
\newblock \bibinfo{title}{V2S attack: building DNN-based voice conversion from
  automatic speaker verification}.
\newblock
\newblock
\showeprint[arxiv]{1908.01454}~[cs.SD]


\bibitem[\protect\citeauthoryear{{Nasr}, {Shokri}, and {Houmansadr}}{{Nasr}
  et~al\mbox{.}}{2019}]%
        {8835245}
\bibfield{author}{\bibinfo{person}{M. {Nasr}}, \bibinfo{person}{R. {Shokri}},
  {and} \bibinfo{person}{A. {Houmansadr}}.} \bibinfo{year}{2019}\natexlab{}.
\newblock \showarticletitle{Comprehensive Privacy Analysis of Deep Learning:
  Passive and Active White-box Inference Attacks against Centralized and
  Federated Learning}. In \bibinfo{booktitle}{\emph{2019 IEEE Symposium on
  Security and Privacy (SP)}}. \bibinfo{pages}{739--753}.
\newblock
\urldef\tempurl%
\url{https://doi.org/10.1109/SP.2019.00065}
\showDOI{\tempurl}


\bibitem[\protect\citeauthoryear{Nautsch, Patino, Tomashenko, Yamagishi, Noé,
  Bonastre, Todisco, and Evans}{Nautsch et~al\mbox{.}}{2020}]%
        {Nautsch_2020}
\bibfield{author}{\bibinfo{person}{Andreas Nautsch}, \bibinfo{person}{Jose
  Patino}, \bibinfo{person}{N. Tomashenko}, \bibinfo{person}{Junichi
  Yamagishi}, \bibinfo{person}{Paul-Gauthier Noé},
  \bibinfo{person}{Jean-François Bonastre}, \bibinfo{person}{Massimiliano
  Todisco}, {and} \bibinfo{person}{Nicholas Evans}.}
  \bibinfo{year}{2020}\natexlab{}.
\newblock \showarticletitle{The Privacy ZEBRA: Zero Evidence Biometric
  Recognition Assessment}.
\newblock \bibinfo{journal}{\emph{Interspeech 2020}} (\bibinfo{date}{Oct}
  \bibinfo{year}{2020}).
\newblock
\urldef\tempurl%
\url{https://doi.org/10.21437/interspeech.2020-1815}
\showDOI{\tempurl}


\bibitem[\protect\citeauthoryear{Oord, Dieleman, Zen, Simonyan, Vinyals,
  Graves, Kalchbrenner, Senior, and Kavukcuoglu}{Oord et~al\mbox{.}}{2016}]%
        {oord2016wavenet}
\bibfield{author}{\bibinfo{person}{Aaron van~den Oord}, \bibinfo{person}{Sander
  Dieleman}, \bibinfo{person}{Heiga Zen}, \bibinfo{person}{Karen Simonyan},
  \bibinfo{person}{Oriol Vinyals}, \bibinfo{person}{Alex Graves},
  \bibinfo{person}{Nal Kalchbrenner}, \bibinfo{person}{Andrew Senior}, {and}
  \bibinfo{person}{Koray Kavukcuoglu}.} \bibinfo{year}{2016}\natexlab{}.
\newblock \showarticletitle{Wavenet: A generative model for raw audio}.
\newblock \bibinfo{journal}{\emph{arXiv preprint arXiv:1609.03499}}
  (\bibinfo{year}{2016}).
\newblock


\bibitem[\protect\citeauthoryear{Oprisanu, Gascon, and De~Cristofaro}{Oprisanu
  et~al\mbox{.}}{[n.d.]}]%
        {oprisanuevaluating}
\bibfield{author}{\bibinfo{person}{Bristena Oprisanu}, \bibinfo{person}{Adria
  Gascon}, {and} \bibinfo{person}{Emiliano De~Cristofaro}.}
  \bibinfo{year}{[n.d.]}\natexlab{}.
\newblock \showarticletitle{Evaluating Privacy-Preserving Generative Models in
  the Wild Technical Report}.
\newblock  (\bibinfo{year}{[n.\,d.]}).
\newblock


\bibitem[\protect\citeauthoryear{Papernot, McDaniel, and Goodfellow}{Papernot
  et~al\mbox{.}}{2016}]%
        {papernot2016transferability}
\bibfield{author}{\bibinfo{person}{Nicolas Papernot}, \bibinfo{person}{Patrick
  McDaniel}, {and} \bibinfo{person}{Ian Goodfellow}.}
  \bibinfo{year}{2016}\natexlab{}.
\newblock \showarticletitle{Transferability in machine learning: from phenomena
  to black-box attacks using adversarial samples}.
\newblock \bibinfo{journal}{\emph{arXiv preprint arXiv:1605.07277}}
  (\bibinfo{year}{2016}).
\newblock


\bibitem[\protect\citeauthoryear{Paulik, Seigel, Mason, Telaar, Kluivers, van
  Dalen, Lau, Carlson, Granqvist, Vandevelde, et~al\mbox{.}}{Paulik
  et~al\mbox{.}}{2021}]%
        {paulik2021federated}
\bibfield{author}{\bibinfo{person}{Matthias Paulik}, \bibinfo{person}{Matt
  Seigel}, \bibinfo{person}{Henry Mason}, \bibinfo{person}{Dominic Telaar},
  \bibinfo{person}{Joris Kluivers}, \bibinfo{person}{Rogier van Dalen},
  \bibinfo{person}{Chi~Wai Lau}, \bibinfo{person}{Luke Carlson},
  \bibinfo{person}{Filip Granqvist}, \bibinfo{person}{Chris Vandevelde},
  {et~al\mbox{.}}} \bibinfo{year}{2021}\natexlab{}.
\newblock \showarticletitle{Federated Evaluation and Tuning for On-Device
  Personalization: System Design \& Applications}.
\newblock \bibinfo{journal}{\emph{arXiv preprint arXiv:2102.08503}}
  (\bibinfo{year}{2021}).
\newblock


\bibitem[\protect\citeauthoryear{Peri, Li, Somandepalli, Jati, and
  Narayanan}{Peri et~al\mbox{.}}{2020}]%
        {peri2020empirical}
\bibfield{author}{\bibinfo{person}{Raghuveer Peri}, \bibinfo{person}{Haoqi Li},
  \bibinfo{person}{Krishna Somandepalli}, \bibinfo{person}{Arindam Jati}, {and}
  \bibinfo{person}{Shrikanth Narayanan}.} \bibinfo{year}{2020}\natexlab{}.
\newblock \showarticletitle{An empirical analysis of information encoded in
  disentangled neural speaker representations}. In
  \bibinfo{booktitle}{\emph{Proceedings of Odyssey}} (Tokyo, Japan).
\newblock


\bibitem[\protect\citeauthoryear{Povey, Ghoshal, Boulianne, Burget, Glembek,
  Goel, Hannemann, Motlicek, Qian, Schwarz, et~al\mbox{.}}{Povey
  et~al\mbox{.}}{2011}]%
        {povey2011kaldi}
\bibfield{author}{\bibinfo{person}{Daniel Povey}, \bibinfo{person}{Arnab
  Ghoshal}, \bibinfo{person}{Gilles Boulianne}, \bibinfo{person}{Lukas Burget},
  \bibinfo{person}{Ondrej Glembek}, \bibinfo{person}{Nagendra Goel},
  \bibinfo{person}{Mirko Hannemann}, \bibinfo{person}{Petr Motlicek},
  \bibinfo{person}{Yanmin Qian}, \bibinfo{person}{Petr Schwarz},
  {et~al\mbox{.}}} \bibinfo{year}{2011}\natexlab{}.
\newblock \showarticletitle{The Kaldi speech recognition toolkit}. In
  \bibinfo{booktitle}{\emph{IEEE 2011 workshop on automatic speech recognition
  and understanding}}. IEEE Signal Processing Society.
\newblock


\bibitem[\protect\citeauthoryear{Qian, Du, Hou, Chen, Jung, and Li}{Qian
  et~al\mbox{.}}{2018}]%
        {Hidebehind_2018}
\bibfield{author}{\bibinfo{person}{Jianwei Qian}, \bibinfo{person}{Haohua Du},
  \bibinfo{person}{Jiahui Hou}, \bibinfo{person}{Linlin Chen},
  \bibinfo{person}{Taeho Jung}, {and} \bibinfo{person}{Xiang-Yang Li}.}
  \bibinfo{year}{2018}\natexlab{}.
\newblock \showarticletitle{Hidebehind: Enjoy Voice Input with Voiceprint
  Unclonability and Anonymity}. In \bibinfo{booktitle}{\emph{Proceedings of the
  16th ACM Conference on Embedded Networked Sensor Systems}} (Shenzhen, China).
  \bibinfo{publisher}{Association for Computing Machinery},
  \bibinfo{pages}{82–94}.
\newblock
\urldef\tempurl%
\url{https://doi.org/10.1145/3274783.3274855}
\showDOI{\tempurl}


\bibitem[\protect\citeauthoryear{Qin, Carlini, Cottrell, Goodfellow, and
  Raffel}{Qin et~al\mbox{.}}{2019}]%
        {qin2019imperceptible}
\bibfield{author}{\bibinfo{person}{Yao Qin}, \bibinfo{person}{Nicholas
  Carlini}, \bibinfo{person}{Garrison Cottrell}, \bibinfo{person}{Ian
  Goodfellow}, {and} \bibinfo{person}{Colin Raffel}.}
  \bibinfo{year}{2019}\natexlab{}.
\newblock \showarticletitle{Imperceptible, robust, and targeted adversarial
  examples for automatic speech recognition}. In
  \bibinfo{booktitle}{\emph{International conference on machine learning}}.
  PMLR, \bibinfo{pages}{5231--5240}.
\newblock


\bibitem[\protect\citeauthoryear{Reynolds, Quatieri, and Dunn}{Reynolds
  et~al\mbox{.}}{2000}]%
        {reynolds2000speaker}
\bibfield{author}{\bibinfo{person}{Douglas~A Reynolds},
  \bibinfo{person}{Thomas~F Quatieri}, {and} \bibinfo{person}{Robert~B Dunn}.}
  \bibinfo{year}{2000}\natexlab{}.
\newblock \showarticletitle{Speaker verification using adapted Gaussian mixture
  models}.
\newblock \bibinfo{journal}{\emph{Digital signal processing}}
  \bibinfo{volume}{10}, \bibinfo{number}{1-3} (\bibinfo{year}{2000}),
  \bibinfo{pages}{19--41}.
\newblock


\bibitem[\protect\citeauthoryear{Rogers, Bater, He, Machanavajjhala, Suresh,
  and Wang}{Rogers et~al\mbox{.}}{2019}]%
        {rogers2019privacy}
\bibfield{author}{\bibinfo{person}{Jennie Rogers}, \bibinfo{person}{Johes
  Bater}, \bibinfo{person}{Xi He}, \bibinfo{person}{Ashwin Machanavajjhala},
  \bibinfo{person}{Madhav Suresh}, {and} \bibinfo{person}{Xiao Wang}.}
  \bibinfo{year}{2019}\natexlab{}.
\newblock \showarticletitle{Privacy changes everything}.
\newblock In \bibinfo{booktitle}{\emph{Heterogeneous data management,
  polystores, and analytics for healthcare}}. \bibinfo{publisher}{Springer},
  \bibinfo{pages}{96--111}.
\newblock


\bibitem[\protect\citeauthoryear{Sch{\"o}nherr, Kohls, Zeiler, Holz, and
  Kolossa}{Sch{\"o}nherr et~al\mbox{.}}{2018}]%
        {schonherr2018adversarial}
\bibfield{author}{\bibinfo{person}{Lea Sch{\"o}nherr},
  \bibinfo{person}{Katharina Kohls}, \bibinfo{person}{Steffen Zeiler},
  \bibinfo{person}{Thorsten Holz}, {and} \bibinfo{person}{Dorothea Kolossa}.}
  \bibinfo{year}{2018}\natexlab{}.
\newblock \showarticletitle{Adversarial attacks against automatic speech
  recognition systems via psychoacoustic hiding}.
\newblock \bibinfo{journal}{\emph{arXiv preprint arXiv:1808.05665}}
  (\bibinfo{year}{2018}).
\newblock


\bibitem[\protect\citeauthoryear{Schuller and Batliner}{Schuller and
  Batliner}{1988}]%
        {schuller1988emotion}
\bibfield{author}{\bibinfo{person}{Bj{\"o}rn~W Schuller} {and}
  \bibinfo{person}{Anton~M Batliner}.} \bibinfo{year}{1988}\natexlab{}.
\newblock \showarticletitle{EMOTION, AFFECT AND PERSONALITY IN SPEECH AND
  LANGUAGE PROCESSING}.
\newblock  (\bibinfo{year}{1988}).
\newblock


\bibitem[\protect\citeauthoryear{Sim, Zadrazil, and Beaufays}{Sim
  et~al\mbox{.}}{2019}]%
        {sim2019investigation}
\bibfield{author}{\bibinfo{person}{Khe~Chai Sim}, \bibinfo{person}{Petr
  Zadrazil}, {and} \bibinfo{person}{Fran{\c{c}}oise Beaufays}.}
  \bibinfo{year}{2019}\natexlab{}.
\newblock \showarticletitle{An investigation into on-device personalization of
  end-to-end automatic speech recognition models}.
\newblock \bibinfo{journal}{\emph{arXiv preprint arXiv:1909.06678}}
  (\bibinfo{year}{2019}).
\newblock


\bibitem[\protect\citeauthoryear{{Sisman}, {Yamagishi}, {King}, and
  {Li}}{{Sisman} et~al\mbox{.}}{2021}]%
        {9262021}
\bibfield{author}{\bibinfo{person}{B. {Sisman}}, \bibinfo{person}{J.
  {Yamagishi}}, \bibinfo{person}{S. {King}}, {and} \bibinfo{person}{H. {Li}}.}
  \bibinfo{year}{2021}\natexlab{}.
\newblock \showarticletitle{An Overview of Voice Conversion and Its Challenges:
  From Statistical Modeling to Deep Learning}.
\newblock \bibinfo{journal}{\emph{IEEE/ACM Transactions on Audio, Speech, and
  Language Processing}}  \bibinfo{volume}{29} (\bibinfo{year}{2021}),
  \bibinfo{pages}{132--157}.
\newblock
\urldef\tempurl%
\url{https://doi.org/10.1109/TASLP.2020.3038524}
\showDOI{\tempurl}


\bibitem[\protect\citeauthoryear{Snyder, Garcia-Romero, Sell, Povey, and
  Khudanpur}{Snyder et~al\mbox{.}}{2018}]%
        {snyder2018x}
\bibfield{author}{\bibinfo{person}{David Snyder}, \bibinfo{person}{Daniel
  Garcia-Romero}, \bibinfo{person}{Gregory Sell}, \bibinfo{person}{Daniel
  Povey}, {and} \bibinfo{person}{Sanjeev Khudanpur}.}
  \bibinfo{year}{2018}\natexlab{}.
\newblock \showarticletitle{X-vectors: Robust dnn embeddings for speaker
  recognition}. In \bibinfo{booktitle}{\emph{2018 IEEE International Conference
  on Acoustics, Speech and Signal Processing (ICASSP)}}. IEEE,
  \bibinfo{pages}{5329--5333}.
\newblock


\bibitem[\protect\citeauthoryear{Song and Shmatikov}{Song and
  Shmatikov}{2019}]%
        {song2019auditing}
\bibfield{author}{\bibinfo{person}{Congzheng Song} {and}
  \bibinfo{person}{Vitaly Shmatikov}.} \bibinfo{year}{2019}\natexlab{}.
\newblock \showarticletitle{Auditing data provenance in text-generation
  models}. In \bibinfo{booktitle}{\emph{Proceedings of the 25th ACM SIGKDD
  International Conference on Knowledge Discovery \& Data Mining}}.
  \bibinfo{pages}{196--206}.
\newblock


\bibitem[\protect\citeauthoryear{Song and Shmatikov}{Song and
  Shmatikov}{2020}]%
        {song2020overlearning}
\bibfield{author}{\bibinfo{person}{Congzheng Song} {and}
  \bibinfo{person}{Vitaly Shmatikov}.} \bibinfo{year}{2020}\natexlab{}.
\newblock \bibinfo{title}{Overlearning Reveals Sensitive Attributes}.
\newblock
\newblock
\showeprint[arxiv]{1905.11742}~[cs.LG]


\bibitem[\protect\citeauthoryear{Song, Shokri, and Mittal}{Song
  et~al\mbox{.}}{2019}]%
        {song2019privacy}
\bibfield{author}{\bibinfo{person}{Liwei Song}, \bibinfo{person}{Reza Shokri},
  {and} \bibinfo{person}{Prateek Mittal}.} \bibinfo{year}{2019}\natexlab{}.
\newblock \showarticletitle{Privacy risks of securing machine learning models
  against adversarial examples}. In \bibinfo{booktitle}{\emph{Proceedings of
  the 2019 ACM SIGSAC Conference on Computer and Communications Security}}.
  \bibinfo{pages}{241--257}.
\newblock


\bibitem[\protect\citeauthoryear{Srivastava, Bellet, Tommasi, and
  Vincent}{Srivastava et~al\mbox{.}}{2019a}]%
        {Srivastava_2019}
\bibfield{author}{\bibinfo{person}{Brij Mohan~Lal Srivastava},
  \bibinfo{person}{Aurélien Bellet}, \bibinfo{person}{Marc Tommasi}, {and}
  \bibinfo{person}{Emmanuel Vincent}.} \bibinfo{year}{2019}\natexlab{a}.
\newblock \showarticletitle{Privacy-Preserving Adversarial Representation
  Learning in ASR: Reality or Illusion?}
\newblock \bibinfo{journal}{\emph{Interspeech 2019}} (\bibinfo{date}{Sep}
  \bibinfo{year}{2019}).
\newblock
\urldef\tempurl%
\url{https://doi.org/10.21437/interspeech.2019-2415}
\showDOI{\tempurl}


\bibitem[\protect\citeauthoryear{Srivastava, Bellet, Tommasi, and
  Vincent}{Srivastava et~al\mbox{.}}{2019b}]%
        {srivastava2019privacy}
\bibfield{author}{\bibinfo{person}{Brij Mohan~Lal Srivastava},
  \bibinfo{person}{Aur{\'e}lien Bellet}, \bibinfo{person}{Marc Tommasi}, {and}
  \bibinfo{person}{Emmanuel Vincent}.} \bibinfo{year}{2019}\natexlab{b}.
\newblock \showarticletitle{Privacy-preserving adversarial representation
  learning in ASR: Reality or illusion?}
\newblock \bibinfo{journal}{\emph{arXiv preprint arXiv:1911.04913}}
  (\bibinfo{year}{2019}).
\newblock


\bibitem[\protect\citeauthoryear{Srivastava, Tomashenko, Wang, Vincent,
  Yamagishi, Maouche, Bellet, and Tommasi}{Srivastava et~al\mbox{.}}{2020}]%
        {srivastava2020design}
\bibfield{author}{\bibinfo{person}{Brij Mohan~Lal Srivastava},
  \bibinfo{person}{Natalia Tomashenko}, \bibinfo{person}{Xin Wang},
  \bibinfo{person}{Emmanuel Vincent}, \bibinfo{person}{Junichi Yamagishi},
  \bibinfo{person}{Mohamed Maouche}, \bibinfo{person}{Aurélien Bellet}, {and}
  \bibinfo{person}{Marc Tommasi}.} \bibinfo{year}{2020}\natexlab{}.
\newblock \bibinfo{title}{Design Choices for X-vector Based Speaker
  Anonymization}.
\newblock
\newblock
\showeprint[arxiv]{2005.08601}~[eess.AS]


\bibitem[\protect\citeauthoryear{Taori, Kamsetty, Chu, and Vemuri}{Taori
  et~al\mbox{.}}{2019}]%
        {taori2019targeted}
\bibfield{author}{\bibinfo{person}{Rohan Taori}, \bibinfo{person}{Amog
  Kamsetty}, \bibinfo{person}{Brenton Chu}, {and} \bibinfo{person}{Nikita
  Vemuri}.} \bibinfo{year}{2019}\natexlab{}.
\newblock \showarticletitle{Targeted adversarial examples for black box audio
  systems}. In \bibinfo{booktitle}{\emph{2019 IEEE Security and Privacy
  Workshops (SPW)}}. IEEE, \bibinfo{pages}{15--20}.
\newblock


\bibitem[\protect\citeauthoryear{Tian, Cai, He, and Liu}{Tian
  et~al\mbox{.}}{2015}]%
        {tian2015investigation}
\bibfield{author}{\bibinfo{person}{Yao Tian}, \bibinfo{person}{Meng Cai},
  \bibinfo{person}{Liang He}, {and} \bibinfo{person}{Jia Liu}.}
  \bibinfo{year}{2015}\natexlab{}.
\newblock \showarticletitle{Investigation of bottleneck features and
  multilingual deep neural networks for speaker verification}. In
  \bibinfo{booktitle}{\emph{Sixteenth Annual Conference of the International
  Speech Communication Association}}.
\newblock


\bibitem[\protect\citeauthoryear{Tomashenko, Srivastava, Wang, Vincent,
  Nautsch, Yamagishi, Evans, Patino, Bonastre, No{\'e},
  et~al\mbox{.}}{Tomashenko et~al\mbox{.}}{2020}]%
        {tomashenko2020introducing}
\bibfield{author}{\bibinfo{person}{Natalia Tomashenko}, \bibinfo{person}{Brij
  Mohan~Lal Srivastava}, \bibinfo{person}{Xin Wang}, \bibinfo{person}{Emmanuel
  Vincent}, \bibinfo{person}{Andreas Nautsch}, \bibinfo{person}{Junichi
  Yamagishi}, \bibinfo{person}{Nicholas Evans}, \bibinfo{person}{Jose Patino},
  \bibinfo{person}{Jean-Fran{\c{c}}ois Bonastre},
  \bibinfo{person}{Paul-Gauthier No{\'e}}, {et~al\mbox{.}}}
  \bibinfo{year}{2020}\natexlab{}.
\newblock \showarticletitle{Introducing the VoicePrivacy initiative}.
\newblock \bibinfo{journal}{\emph{arXiv preprint arXiv:2005.01387}}
  (\bibinfo{year}{2020}).
\newblock


\bibitem[\protect\citeauthoryear{{Vaidya} and {Sherr}}{{Vaidya} and
  {Sherr}}{2019}]%
        {8844600}
\bibfield{author}{\bibinfo{person}{T. {Vaidya}} {and} \bibinfo{person}{M.
  {Sherr}}.} \bibinfo{year}{2019}\natexlab{}.
\newblock \showarticletitle{You Talk Too Much: Limiting Privacy Exposure Via
  Voice Input}. In \bibinfo{booktitle}{\emph{2019 IEEE Security and Privacy
  Workshops (SPW)}}. \bibinfo{pages}{84--91}.
\newblock
\urldef\tempurl%
\url{https://doi.org/10.1109/SPW.2019.00026}
\showDOI{\tempurl}


\bibitem[\protect\citeauthoryear{Variani, Lei, McDermott, Moreno, and
  Gonzalez-Dominguez}{Variani et~al\mbox{.}}{2014}]%
        {variani2014deep}
\bibfield{author}{\bibinfo{person}{Ehsan Variani}, \bibinfo{person}{Xin Lei},
  \bibinfo{person}{Erik McDermott}, \bibinfo{person}{Ignacio~Lopez Moreno},
  {and} \bibinfo{person}{Javier Gonzalez-Dominguez}.}
  \bibinfo{year}{2014}\natexlab{}.
\newblock \showarticletitle{Deep neural networks for small footprint
  text-dependent speaker verification}. In \bibinfo{booktitle}{\emph{2014 IEEE
  International Conference on Acoustics, Speech and Signal Processing
  (ICASSP)}}. IEEE, \bibinfo{pages}{4052--4056}.
\newblock


\bibitem[\protect\citeauthoryear{Veaux, Yamagishi, and King}{Veaux
  et~al\mbox{.}}{2013}]%
        {veaux2013towards}
\bibfield{author}{\bibinfo{person}{Christophe Veaux}, \bibinfo{person}{Junichi
  Yamagishi}, {and} \bibinfo{person}{Simon King}.}
  \bibinfo{year}{2013}\natexlab{}.
\newblock \showarticletitle{Towards personalised synthesised voices for
  individuals with vocal disabilities: Voice banking and reconstruction}. In
  \bibinfo{booktitle}{\emph{Proceedings of the Fourth Workshop on Speech and
  Language Processing for Assistive Technologies}}. \bibinfo{pages}{107--111}.
\newblock


\bibitem[\protect\citeauthoryear{Villalba, Chen, Snyder, Garcia-Romero, McCree,
  Sell, Borgstrom, Garc{\'\i}a-Perera, Richardson, Dehak,
  et~al\mbox{.}}{Villalba et~al\mbox{.}}{2020a}]%
        {villalba2020state}
\bibfield{author}{\bibinfo{person}{Jes{\'u}s Villalba}, \bibinfo{person}{Nanxin
  Chen}, \bibinfo{person}{David Snyder}, \bibinfo{person}{Daniel
  Garcia-Romero}, \bibinfo{person}{Alan McCree}, \bibinfo{person}{Gregory
  Sell}, \bibinfo{person}{Jonas Borgstrom}, \bibinfo{person}{Leibny~Paola
  Garc{\'\i}a-Perera}, \bibinfo{person}{Fred Richardson},
  \bibinfo{person}{R{\'e}da Dehak}, {et~al\mbox{.}}}
  \bibinfo{year}{2020}\natexlab{a}.
\newblock \showarticletitle{State-of-the-art speaker recognition with neural
  network embeddings in NIST SRE18 and speakers in the wild evaluations}.
\newblock \bibinfo{journal}{\emph{Computer Speech \& Language}}
  \bibinfo{volume}{60} (\bibinfo{year}{2020}), \bibinfo{pages}{101026}.
\newblock


\bibitem[\protect\citeauthoryear{Villalba, Zhang, and Dehak}{Villalba
  et~al\mbox{.}}{2020b}]%
        {villalba2020x}
\bibfield{author}{\bibinfo{person}{Jes{\'u}s Villalba}, \bibinfo{person}{Yuekai
  Zhang}, {and} \bibinfo{person}{Najim Dehak}.}
  \bibinfo{year}{2020}\natexlab{b}.
\newblock \showarticletitle{x-Vectors Meet Adversarial Attacks: Benchmarking
  Adversarial Robustness in Speaker Verification}.
\newblock \bibinfo{journal}{\emph{Proc. Interspeech 2020}}
  (\bibinfo{year}{2020}), \bibinfo{pages}{4233--4237}.
\newblock


\bibitem[\protect\citeauthoryear{Wang, Yamagishi, Todisco, Delgado, Nautsch,
  Evans, Sahidullah, Vestman, Kinnunen, Lee, Juvela, Alku, Peng, Hwang, Tsao,
  Wang, Maguer, Becker, Henderson, Clark, Zhang, Wang, Jia, Onuma, Mushika,
  Kaneda, Jiang, Liu, Wu, Huang, Toda, Tanaka, Kameoka, Steiner, Matrouf,
  Bonastre, Govender, Ronanki, Zhang, and Ling}{Wang et~al\mbox{.}}{2020}]%
        {wang2020asvspoof}
\bibfield{author}{\bibinfo{person}{Xin Wang}, \bibinfo{person}{Junichi
  Yamagishi}, \bibinfo{person}{Massimiliano Todisco}, \bibinfo{person}{Hector
  Delgado}, \bibinfo{person}{Andreas Nautsch}, \bibinfo{person}{Nicholas
  Evans}, \bibinfo{person}{Md Sahidullah}, \bibinfo{person}{Ville Vestman},
  \bibinfo{person}{Tomi Kinnunen}, \bibinfo{person}{Kong~Aik Lee},
  \bibinfo{person}{Lauri Juvela}, \bibinfo{person}{Paavo Alku},
  \bibinfo{person}{Yu-Huai Peng}, \bibinfo{person}{Hsin-Te Hwang},
  \bibinfo{person}{Yu Tsao}, \bibinfo{person}{Hsin-Min Wang},
  \bibinfo{person}{Sebastien~Le Maguer}, \bibinfo{person}{Markus Becker},
  \bibinfo{person}{Fergus Henderson}, \bibinfo{person}{Rob Clark},
  \bibinfo{person}{Yu Zhang}, \bibinfo{person}{Quan Wang}, \bibinfo{person}{Ye
  Jia}, \bibinfo{person}{Kai Onuma}, \bibinfo{person}{Koji Mushika},
  \bibinfo{person}{Takashi Kaneda}, \bibinfo{person}{Yuan Jiang},
  \bibinfo{person}{Li-Juan Liu}, \bibinfo{person}{Yi-Chiao Wu},
  \bibinfo{person}{Wen-Chin Huang}, \bibinfo{person}{Tomoki Toda},
  \bibinfo{person}{Kou Tanaka}, \bibinfo{person}{Hirokazu Kameoka},
  \bibinfo{person}{Ingmar Steiner}, \bibinfo{person}{Driss Matrouf},
  \bibinfo{person}{Jean-Francois Bonastre}, \bibinfo{person}{Avashna Govender},
  \bibinfo{person}{Srikanth Ronanki}, \bibinfo{person}{Jing-Xuan Zhang}, {and}
  \bibinfo{person}{Zhen-Hua Ling}.} \bibinfo{year}{2020}\natexlab{}.
\newblock \bibinfo{title}{ASVspoof 2019: A large-scale public database of
  synthesized, converted and replayed speech}.
\newblock
\newblock
\showeprint[arxiv]{1911.01601}~[eess.AS]


\bibitem[\protect\citeauthoryear{Wester}{Wester}{2010}]%
        {wester2010emime}
\bibfield{author}{\bibinfo{person}{Mirjam Wester}.}
  \bibinfo{year}{2010}\natexlab{}.
\newblock \bibinfo{booktitle}{\emph{The EMIME bilingual database}}.
\newblock \bibinfo{type}{{T}echnical {R}eport}. \bibinfo{institution}{The
  University of Edinburgh}.
\newblock


\bibitem[\protect\citeauthoryear{Wu, Liu, and Lee}{Wu et~al\mbox{.}}{2020b}]%
        {wu2020defense}
\bibfield{author}{\bibinfo{person}{Haibin Wu}, \bibinfo{person}{Andy~T Liu},
  {and} \bibinfo{person}{Hung-yi Lee}.} \bibinfo{year}{2020}\natexlab{b}.
\newblock \showarticletitle{Defense for black-box attacks on anti-spoofing
  models by self-supervised learning}.
\newblock \bibinfo{journal}{\emph{arXiv preprint arXiv:2006.03214}}
  (\bibinfo{year}{2020}).
\newblock


\bibitem[\protect\citeauthoryear{Wu, Das, Yang, and Li}{Wu
  et~al\mbox{.}}{2020a}]%
        {wu2020light}
\bibfield{author}{\bibinfo{person}{Zhenzong Wu}, \bibinfo{person}{Rohan~Kumar
  Das}, \bibinfo{person}{Jichen Yang}, {and} \bibinfo{person}{Haizhou Li}.}
  \bibinfo{year}{2020}\natexlab{a}.
\newblock \showarticletitle{Light convolutional neural network with feature
  genuinization for detection of synthetic speech attacks}.
\newblock \bibinfo{journal}{\emph{arXiv preprint arXiv:2009.09637}}
  (\bibinfo{year}{2020}).
\newblock


\bibitem[\protect\citeauthoryear{Wu, Evans, Kinnunen, Yamagishi, Alegre, and
  Li}{Wu et~al\mbox{.}}{2015}]%
        {13}
\bibfield{author}{\bibinfo{person}{Zhizheng Wu}, \bibinfo{person}{Nicholas
  Evans}, \bibinfo{person}{Tomi Kinnunen}, \bibinfo{person}{Junichi Yamagishi},
  \bibinfo{person}{Federico Alegre}, {and} \bibinfo{person}{Haizhou Li}.}
  \bibinfo{year}{2015}\natexlab{}.
\newblock \showarticletitle{Spoofing and countermeasures for speaker
  verification: A survey}.
\newblock \bibinfo{journal}{\emph{speech communication}}
  (\bibinfo{year}{2015}), \bibinfo{pages}{130--153}.
\newblock


\bibitem[\protect\citeauthoryear{Wu and Li}{Wu and Li}{2014}]%
        {wu2014voice}
\bibfield{author}{\bibinfo{person}{Zhizheng Wu} {and} \bibinfo{person}{Haizhou
  Li}.} \bibinfo{year}{2014}\natexlab{}.
\newblock \showarticletitle{Voice conversion versus speaker verification: an
  overview}.
\newblock \bibinfo{journal}{\emph{APSIPA Transactions on Signal and Information
  Processing}}  \bibinfo{volume}{3} (\bibinfo{year}{2014}).
\newblock


\bibitem[\protect\citeauthoryear{y.~{Huang}, {Lin}, y.~{Lee}, and
  s.~{Lee}}{y.~{Huang} et~al\mbox{.}}{2021}]%
        {9383529}
\bibfield{author}{\bibinfo{person}{C. y. {Huang}}, \bibinfo{person}{Y.~Y.
  {Lin}}, \bibinfo{person}{H. y. {Lee}}, {and} \bibinfo{person}{L. s. {Lee}}.}
  \bibinfo{year}{2021}\natexlab{}.
\newblock \showarticletitle{Defending Your Voice: Adversarial Attack on Voice
  Conversion}. In \bibinfo{booktitle}{\emph{2021 IEEE Spoken Language
  Technology Workshop (SLT)}}. \bibinfo{pages}{552--559}.
\newblock
\urldef\tempurl%
\url{https://doi.org/10.1109/SLT48900.2021.9383529}
\showDOI{\tempurl}


\bibitem[\protect\citeauthoryear{Yamagishi, Veaux, and MacDonald}{Yamagishi
  et~al\mbox{.}}{2019}]%
        {yamagishi2019vctk}
\bibfield{author}{\bibinfo{person}{Junichi Yamagishi},
  \bibinfo{person}{Christophe Veaux}, {and} \bibinfo{person}{Kirsten
  MacDonald}.} \bibinfo{year}{2019}\natexlab{}.
\newblock \bibinfo{title}{{CSTR VCTK Corpus}: English Multi-speaker Corpus for
  {CSTR} Voice Cloning Toolkit (version 0.92)}.
\newblock
\newblock
\urldef\tempurl%
\url{https://doi.org/10.7488/ds/2645}
\showDOI{\tempurl}


\bibitem[\protect\citeauthoryear{Yi, Huang, Tian, Yamagishi, Das, Kinnunen,
  Ling, and Toda}{Yi et~al\mbox{.}}{2020}]%
        {Yi2020}
\bibfield{author}{\bibinfo{person}{Zhao Yi}, \bibinfo{person}{Wen-Chin Huang},
  \bibinfo{person}{Xiaohai Tian}, \bibinfo{person}{Junichi Yamagishi},
  \bibinfo{person}{Rohan~Kumar Das}, \bibinfo{person}{Tomi Kinnunen},
  \bibinfo{person}{Zhen-Hua Ling}, {and} \bibinfo{person}{Tomoki Toda}.}
  \bibinfo{year}{2020}\natexlab{}.
\newblock \showarticletitle{{Voice Conversion Challenge 2020 –- Intra-lingual
  semi-parallel and cross-lingual voice conversion –-}}. In
  \bibinfo{booktitle}{\emph{Proc. Joint Workshop for the Blizzard Challenge and
  Voice Conversion Challenge 2020}}. \bibinfo{pages}{80--98}.
\newblock
\urldef\tempurl%
\url{https://doi.org/10.21437/VCC_BC.2020-14}
\showDOI{\tempurl}


\bibitem[\protect\citeauthoryear{{Yoo}, {Lee}, {Leem}, {Oh}, {Ko}, and
  {Yook}}{{Yoo} et~al\mbox{.}}{2020}]%
        {9247219}
\bibfield{author}{\bibinfo{person}{I.~C. {Yoo}}, \bibinfo{person}{K. {Lee}},
  \bibinfo{person}{S. {Leem}}, \bibinfo{person}{H. {Oh}}, \bibinfo{person}{B.
  {Ko}}, {and} \bibinfo{person}{D. {Yook}}.} \bibinfo{year}{2020}\natexlab{}.
\newblock \showarticletitle{Speaker Anonymization for Personal Information
  Protection Using Voice Conversion Techniques}.
\newblock \bibinfo{journal}{\emph{IEEE Access}}  \bibinfo{volume}{8}
  (\bibinfo{year}{2020}), \bibinfo{pages}{198637--198645}.
\newblock
\urldef\tempurl%
\url{https://doi.org/10.1109/ACCESS.2020.3035416}
\showDOI{\tempurl}


\bibitem[\protect\citeauthoryear{Yoon, Drumright, and van~der Schaar}{Yoon
  et~al\mbox{.}}{2020}]%
        {9034117}
\bibfield{author}{\bibinfo{person}{Jinsung Yoon}, \bibinfo{person}{Lydia~N.
  Drumright}, {and} \bibinfo{person}{Mihaela van~der Schaar}.}
  \bibinfo{year}{2020}\natexlab{}.
\newblock \showarticletitle{Anonymization Through Data Synthesis Using
  Generative Adversarial Networks (ADS-GAN)}.
\newblock \bibinfo{journal}{\emph{IEEE Journal of Biomedical and Health
  Informatics}} \bibinfo{volume}{24}, \bibinfo{number}{8}
  (\bibinfo{year}{2020}), \bibinfo{pages}{2378--2388}.
\newblock
\urldef\tempurl%
\url{https://doi.org/10.1109/JBHI.2020.2980262}
\showDOI{\tempurl}


\bibitem[\protect\citeauthoryear{Yuan, Chen, Zhao, Long, Liu, Chen, Zhang,
  Huang, Wang, and Gunter}{Yuan et~al\mbox{.}}{2018}]%
        {yuan2018commandersong}
\bibfield{author}{\bibinfo{person}{Xuejing Yuan}, \bibinfo{person}{Yuxuan
  Chen}, \bibinfo{person}{Yue Zhao}, \bibinfo{person}{Yunhui Long},
  \bibinfo{person}{Xiaokang Liu}, \bibinfo{person}{Kai Chen},
  \bibinfo{person}{Shengzhi Zhang}, \bibinfo{person}{Heqing Huang},
  \bibinfo{person}{Xiaofeng Wang}, {and} \bibinfo{person}{Carl~A Gunter}.}
  \bibinfo{year}{2018}\natexlab{}.
\newblock \showarticletitle{Commandersong: A systematic approach for practical
  adversarial voice recognition}. In \bibinfo{booktitle}{\emph{27th
  $\{$USENIX$\}$ Security Symposium ($\{$USENIX$\}$ Security 18)}}.
  \bibinfo{pages}{49--64}.
\newblock


\bibitem[\protect\citeauthoryear{Zhang, Shirvanian, Arora, Huang, and Gu}{Zhang
  et~al\mbox{.}}{2021}]%
        {zhang2021practical}
\bibfield{author}{\bibinfo{person}{Yangyong Zhang}, \bibinfo{person}{Maliheh
  Shirvanian}, \bibinfo{person}{Sunpreet~S Arora}, \bibinfo{person}{Jianwei
  Huang}, {and} \bibinfo{person}{Guofei Gu}.} \bibinfo{year}{2021}\natexlab{}.
\newblock \showarticletitle{Practical Speech Re-use Prevention in Voice-driven
  Services}.
\newblock \bibinfo{journal}{\emph{arXiv preprint arXiv:2101.04773}}
  (\bibinfo{year}{2021}).
\newblock


\bibitem[\protect\citeauthoryear{{Zheng}, {Li}, {Sun}, {Zhang}, and
  {Zheng}}{{Zheng} et~al\mbox{.}}{2021}]%
        {9198912}
\bibfield{author}{\bibinfo{person}{L. {Zheng}}, \bibinfo{person}{J. {Li}},
  \bibinfo{person}{M. {Sun}}, \bibinfo{person}{X. {Zhang}}, {and}
  \bibinfo{person}{T.~F. {Zheng}}.} \bibinfo{year}{2021}\natexlab{}.
\newblock \showarticletitle{When Automatic Voice Disguise Meets Automatic
  Speaker Verification}.
\newblock \bibinfo{journal}{\emph{IEEE Transactions on Information Forensics
  and Security}}  \bibinfo{volume}{16} (\bibinfo{year}{2021}),
  \bibinfo{pages}{824--837}.
\newblock
\urldef\tempurl%
\url{https://doi.org/10.1109/TIFS.2020.3023818}
\showDOI{\tempurl}


\bibitem[\protect\citeauthoryear{Zheng and Li}{Zheng and Li}{2017}]%
        {zheng2017robustness}
\bibfield{author}{\bibinfo{person}{Thomas~Fang Zheng} {and}
  \bibinfo{person}{Lantian Li}.} \bibinfo{year}{2017}\natexlab{}.
\newblock \bibinfo{booktitle}{\emph{Robustness-related issues in speaker
  recognition}}.
\newblock \bibinfo{publisher}{Springer}.
\newblock


\end{thebibliography}
